\documentclass[a4paper,12pt]{article}

\usepackage{amsmath,amsfonts,amssymb}
\usepackage{a4wide}
\usepackage{natbib,bm}

\def \tr{\mbox{tr\hskip 1pt}}
\def \div{\mbox{div\hskip 1pt}}
\def \Div{\mbox{Div\hskip 1pt}}
\def \grad{\mbox{grad\hskip 1pt}}
\def \Grad{\mbox{Grad\hskip 1pt}}

\numberwithin{equation}{section}

\begin{document}

%%%%%%%%%%%%%%%%%%%%%%%%%%%%%%%%%%%%%%%%%%

\title{\textbf{On stress-dependent elastic moduli and wave speeds}}

%%%%%%%%%%%%%%%%%%%%%%%%%%%%%%%%%%%%%%%%%%

 \author{{\sc M. Destrade}\\[2pt]
 \textit{School of Mathematics, Statistics and Applied Mathematics},  \\[6pt]
   \textit{National University of Ireland, Galway, University Road, Galway, Ireland}\\[2pt]
   michel.destrade@nuigalway.ie\\[6pt]
   {\small AND}\\[6pt]
  {\sc R.W. Ogden$^*$}\\[2pt]
  \textit{School of Engineering, University of Aberdeen},\\[6pt]
    \textit{King's College, Aberdeen AB24 3UE, United Kingdom}\\[2pt]
    $^*$Corresponding author: r.ogden@abdn.ac.uk
    }

\date{}

%%%%%%%%%%%%%%%%%%%%%%%%%%%%%%%%%%%%%%%%%%
\maketitle
%%%%%%%%%%%%%%%%%%%%%%%%%%%%%%%%%%%%%%%%%%

%%%%%%%%%%%%%%%%%%%%%%%%%%%%%%%%%%%%%%%%%%

\vspace{1cm}

\noindent On the basis of the general nonlinear theory of a hyperelastic material with initial stress, initially without consideration of the origin of the initial stress, we determine explicit expressions for the stress-dependent tensor of incremental elastic moduli.  In considering three special cases of initial stress within the general framework, namely hydrostatic stress, uniaxial stress and planar shear stress, we then elucidate in general form the dependence of various elastic moduli on the initial stress.  In each case the effect of initial stress on the wave speed of homogeneous plane waves is studied and it is shown how various special theories from the earlier literature fit within the general framework.  We then consider the situation in which the initial stress is a pre-stress associated with a finite deformation and, in particular, we discuss the specialization to the second-order theory of elasticity and highlight connections between several classical approaches to the topic, again with special reference to the influence of higher-order terms on the speed of homogeneous plane waves.  Some discrepancies arising in the earlier literature are noted.

\vspace{1cm}

\newpage

%%%%%%%%%%%%%%%%%%%%%%%%%%%%%%%%%%%%%%%%%%

\section{Introduction\label{sec1}}

%%%%%%%%%%%%%%%%%%%%%%%%%%%%%%%%%%%%%%%%%%

Residual stresses in solids, i.e. stresses that are present in the absence of load (body forces and surface tractions) can have a
very significant effect on the mechanical behaviour of the structures in which they reside. This is the case for materials as
diverse as hard engineering and geological materials and soft solids such as gels and biological tissues. Equally, stresses that
are generated due to applied loads, associated with finite deformations and commonly referred to as pre-stresses, have a
significant effect on subsequent material response, leading to very different results compared with the  situation in which there
is no applied load.  Residual stresses and pre-stresses are examples of initial stresses but are different in nature in the sense
that residual stresses are necessarily non-homogeneous while pre-stresses may be either homogeneous or non-homogeneous. Also,
pre-stresses are usually associated with an elastic pre-deformation, while residual stresses can result from
processes such as manufacturing, plastic deformation, growth and remodelling, for example. In either case it is important to be able to
analyze the effect of the initial stress on the properties of the material and on its mechanical response. In this paper we are
concerned primarily with the effect of initial stress on the material properties in general, and on elastic `constants' in some
specific cases, as well as its effect on the speeds of propagation of homogeneous plane waves. The study is conducted with a view
to the non-destructive evaluation of initially stressed solids, which are ubiquitous in Nature and Engineering. For this purpose
the initial stress is included in the constitutive description of the material without reference (initially) to any finite
deformation with which it may be associated, and we emphasize that in general the initial stress appears in a highly nonlinear
form.

The origins of an elasticity theory including initial stress can be traced back as far as to the works of  \cite{Cauc29},
according to  \cite{True66}. Notable early   contributors to the subject include \cite{Poin92},  \cite{Hada03}, \cite{Rayl06},
\cite{Bril25}, and \cite{Love27}.

In the context of the modern linear theory of elasticity, the effect of initial stress was first examined in the
work of Biot for static problems \citep{Biot39} and also for wave
propagation problems \citep{Biot40}; see also his monograph  \citep{Biot65}.
Here, for later reference, we write the components of Biot's elasticity tensor as $\mathcal{B}_{piqj}$ with respect
to a Cartesian coordinate system $(x_1,x_2,x_3)$.
In general these enjoy the minor symmetries
\begin{equation}
\mathcal{B}_{piqj}=\mathcal{B}_{ipqj}=\mathcal{B}_{pijq},\label{Biot-conds1}
\end{equation}
but when there exists a strain-energy function there is also the major symmetry connection
\begin{equation}
\mathcal{B}_{piqj}-\mathcal{B}_{qjpi}=\delta_{ip}\tau_{jq}
-\delta_{jq}\tau_{ip},\label{Biot-conds2}
\end{equation}
where $\tau_{ij}$ are the components of the initial Cauchy stress and $\delta_{ij}$ is the Kronecker delta.  For infinitesimal strains $e_{ij}$, with
$e_{ij}=(u_{i,j}+u_{j,i})/2$, where $u_{i,j}=\partial u_i/\partial x_j$ and $u_i$ are the components of the displacement vector,
the associated Cauchy stress, additional to the initial stress, is given by $\sigma_{pi}=\mathcal{B}_{piqj}u_{j,q}$. Biot left
the dependence of $\mathcal{B}_{piqj}$ on the initial stress unspecified for most of his theoretical development and he was not
concerned with the source of the initial stress.  In this sense his theory may be referred to as the \emph{general linear theory of
elasticity with initial stress}. Biot was more specific in particular cases, where he considered both isotropic and planar
orthotropic specializations and an initial stress due to hydrostatic pressure, uniaxial compression, or gravity. In particular,
for his isotropic model $\mathcal{B}_{piqj}$ may be written in the form
\begin{equation}
\mathcal{B}_{piqj}=\mu_0(\delta_{ij}\delta_{pq}+\delta_{qi}\delta_{pj})+\lambda_0\delta_{pi}\delta_{qj}
-\delta_{qj}\tau_{pi},\label{biot-iso}
\end{equation}
where $\mu_0$ and $\lambda_0$ are the notations that we shall use in this paper for the classical Lam\'e moduli of
linear isotropic elasticity. This form did not appear explicitly in Biot's work, as far as the authors are aware, but may be deduced from the plane strain expressions in equation (8.31e) of Biot's book \citep{Biot65}.
Biot did in fact acknowledge that in general the elastic response in the presence of initial stress is not isotropic,
but he adopted an isotropic constitutive description for simplicity.
Note that \eqref{biot-iso} satisfies the conditions \eqref{Biot-conds2}.

The works of Biot have formed the basis for many contributions to the literature, particularly relating to wave
propagation problems in the geophysical context, which was the original context in which the theory was developed
by \cite{Biot40}.  Other contributions to the analysis of initial stress, more specifically residual stress, have
appeared in a series of papers by Hoger and co-workers, including \cite{Hoge85}, \cite{Hoge86}, \cite{Hoge93},
\cite{Hoge93b}, \cite{John93}, \cite{Hoge96}, some of which are concerned with the combined effect of finite deformation
and residual stress, and in papers by  \cite{Man87}, \cite{Man98} and \cite{Sara08}, for example.  More recently, a
general theory of nonlinear hyperelasticity for an initially stressed solid has been developed by \cite{Sham11}
and \cite{Ogde11}, the latter being focused on fibre reinforced materials.   We also mention the paper by \cite{Bazant71}, who
detailed connections between several earlier formulations of linear elasticity with initial stress and their implications for the stability of elastic bodies.

We distinguish between the above approach and that concerned with the effect of pre-stress that is associated
with a finite deformation, a subject that has attracted many contributions, primarily concerned with the effect
of the finite deformation on the propagation of small amplitude elastic waves and associated static bifurcation problems.
This is commonly know as the \emph{theory of incremental (or small) deformations superimposed on a finite deformation}. We shall
not discuss this extensive topic in detail but refer to \cite{Ogde84} and \cite{Ogde07}, for example, for pointers to the literature.

A special case of the theory of finite elastic deformations in which the strains are small but the linear theory is no longer
adequate is sometimes referred to as second-order elasticity.  In this theory the strain-energy function is expanded to the third
order in some suitable measure of strain and the stress is second order in the strain. Most commonly it is the Green (or
Green--Lagrange) strain tensor that is used and for an isotropic material the strain energy is expressed in terms of invariants
of the strain. The first such contribution appears to be that of \cite{Bril25}, although this has not always been acknowledged
appropriately. Equivalent formulations were developed later by \cite{Land37}, \cite{Murn37}, \cite{Biot40x}, \cite{Toup61} and
\cite{Haye61}; see also the books by \cite{Bril46}, \cite{Biot65} and \cite{Land86}. There is also a detailed historical
discussion in the volume by \cite{True65}, in particular in section 66 therein. An important objective within these works was to
determine the correction to the speeds of waves due to the nonlinear terms in the stress. In particular, various formulas were
found that highlighted the effect of an initial hydrostatic pressure (associated with a pure dilatation) or uniaxial compression
on the speeds of longitudinal and transverse waves. Of other works dealing with wave speeds based on this weakly nonlinear theory
we mention those of \cite{HuKe53}, who obtained experimental results for longitudinal and transverse wave speeds in polystyrene,
iron, and pyrex for separate initial stresses corresponding to hydrostatic pressure and simple compression and related their results
to formulas based on Murnaghan's second-order theory, and \cite{Thur64}, who obtained expressions for the second-order corrections
to wave speeds in cubic crystals. Papers by \cite{Birc38} and \cite{Tang67} also made use of the second-order
theory but failed to include the second-order constants in their expressions for the wave speeds (note that the second-order
constants are sometimes referred to as third-order constants). We discuss this and other deficiencies of the latter paper in
Sections 6.1, 6.3 and 7.3.

For several researchers it seemed important to show that all elastic materials would behave similarly under an initial stress.
Hence Biot's incremental moduli \eqref{biot-iso} change from one material to another with changes in the values of $\lambda_0$
and $\mu_0$, but the effect of the pre-stress $\bm\tau$ remains the same across all solids. Similarly, according to
\cite{Laza49}, \cite{Love27} showed that under external pressure $P$, the (linear) elastic constants $c_{44}$ and $c_{12}$
(Voigt notation) of a certain class of solid are linked approximately by the `Cauchy relation' $c_{44} = c_{12}-2P$.
This type of behaviour
under pre-stress would in turn lead to a corresponding effect on the shift in speed experienced by an acoustic wave (for instance a
wave should always travel faster in a pressurized isotropic solid than in its unstressed counterpart). However intuitive this
expectation might be, it is not supported by experimental facts, as confirmed by the data shown in Tables \ref{table1} and
\ref{table2}, which show that the wave speed can increase or decrease with pressure, depending on the material.

\begin{table} [h!]
\caption{\textit{Initial variation of the squared wave speeds for several solids under hydrostatic pressure $P$, as collected by
\cite{John94}: $\rho_r$ is the mass density at $P=0$; $v_T$ and $v_L$ are the speeds of the transverse
and longitudinal waves, respectively}}
\begin{center}
\begin{tabular}{l r r}
    \noalign{\smallskip}\hline\\[-1ex]
Solid & $\left. \dfrac{\text{d}}{\text{d}P}\left(\rho_r v_T^2\right) \right|_{P=0}$  &$\left.
\dfrac{\text{d}}{\text{d}P}\left(\rho_r v_L^2\right) \right|_{P=0}$
\\[16pt]
\hline \\[-4pt]
Alumina & 1.12& 4.46 \\
Aluminum & 2.92 & 12.4 \\
Armco-Iron & 5.7 & 9.3 \\
Fused silica & $-1.42$ & $-4.32$ \\
Gold & 0.90 & 6.4 \\
Magnesium & 1.47 & 6.89 \\
Molybdenum & 1.05 & 3.48 \\
Nickel-steel & 1.55 & 2.84 \\
Niobium & 0.29  & 6.18 \\
PMMA & 3.0 & 15.0 \\
Polystryrene & 1.57 & 11.6 \\
Pyrex & $-2.84$ & $-8.6$ \\
Steel (Hecla) & 1.46 & 7.45 \\
Tungsten & 0.70 & 4.58 \\
    \noalign{\smallskip}\hline
\end{tabular}
\end{center}
 \label{table1}
\end{table}

\begin{table} [h!]
\caption{\textit{Initial variation of the speeds of transverse waves for several solids under uniaxial strain $\epsilon$: $v_{12}$, $v_{21}$
are the speeds waves travelling in the direction of tension, and orthogonal  to the direction of tension, respectively; $v_{12}^0$ and $v_{21}^0$ are their values at  $\epsilon=0$.
The first row of data is from experiments on a sample of rail steel \citep{EgBr76}; the other rows are from experiments on soft solids with different compositions \citep{Genn07}}}
\begin{center}
\begin{tabular}{l r  r}
    \noalign{\smallskip}\hline\\[-1ex]
Solid & $\left. \dfrac{\text{d}}{\text{d}\epsilon}\left(\dfrac{v_{12}}{v_{12}^0}\right) \right|_{\epsilon=0}$  & $\left.
\dfrac{\text{d}}{\text{d}\epsilon}\left(\dfrac{v_{21}}{v_{21}^0}\right) \right|_{\epsilon=0}$
\\[16pt]
\hline \\[-4pt]
Rail steel 1 & $-0.15$ & $-1.50$  \\
Agar-Gelatine 1 & 0.84 & $-0.92$  \\
Agar-Gelatine 4 & 4.53 & 2.26  \\
Polyvinyl acetate 1 & 0.71 & $-0.47$  \\
Polyvinyl acetate 3 & 1.69 & 0.35  \\
    \noalign{\smallskip}\hline
\end{tabular}
\end{center}
 \label{table2}
\end{table}

The purpose of the present paper is to draw together and highlight connections between some of the historical results within a
common and fairly general framework based on the development of \cite{Sham11} concerned with the constitutive law of a
hyperelastic material with initial stress.  In particular, we examine how the elasticity tensor depends nonlinearly on initial
stress, with emphasis on the important special cases of hydrostatic initial stress, uniaxial initial stress, and initial shear
stress.

In Section \ref{sec2} we summarize the basic equations for an elastic material for which the strain-energy function depends on an
initial stress as well as the deformation from the initially stressed reference configuration.  In particular, we express the
strain energy as a function of combined invariants of the initial stress and deformation and give expressions for the nominal and
Cauchy stress tensors.  Next, in Section \ref{sec3}, we derive the equations of motion for small displacements from a
homogeneously deformed configuration when the initial stress is uniform, which leads to the need for expressions for the
elasticity tensor.  Such expressions are given in Section \ref{sec4}, but now specialized to the undeformed (but initially
stressed) reference configuration.  Several special cases are considered in which the initial stress is either purely isotropic,
uniaxial, or a planar shear stress.  These formulas are then used in Section \ref{sec5} to define relevant elastic moduli that
depend of the initial stress in question and to make that dependence explicit.  The moduli include stress-dependent Lam\'e moduli
in the case of isotropic initial stress, Poisson's ratios and Young's moduli for uniaxial initial stress, and planar Poisson's
ratios and Young's moduli for planar initial shear stress.

In Section \ref{sec6} the results are applied to infinitesimal wave propagation and related to some known results as special
cases.  Section \ref{sec7} then considers the deformation of an isotropic elastic material from a stress-free reference
configuration in order to make contact with results in the preceding sections by considering, in particular, a pure dilatation
and a deformation corresponding to simple tension.  We then focus on the specialization to second-order elasticity, in which the
strain-energy is approximated as a third-order expansion in the Green strain tensor in order to highlight connections with the
theory of elasticity with initial stress herein and to draw together various contributions from the literature that date back to
the work of \cite{Bril25}, with particular reference to expressions for longitudinal and transverse wave speeds.

%%%%%%%%%%%%%%%%%%%%%%%%%%%%%%%%%%%%%%%%%%

\section{Elasticity in the presence of initial stress\label{sec2}}

%%%%%%%%%%%%%%%%%%%%%%%%%%%%%%%%%%%%%%%%%%

We consider an elastic body that is subject to an initial (Cauchy) stress $\bm\tau$ in some \emph{reference configuration}, which we
denote by $\mathcal{B}_r$.  In the absence of intrinsic couple stresses $\bm\tau$ is symmetric. Let $\mathbf{X}$ be the position
vector of a material point in  $\mathcal{B}_r$ and let $\Grad$ and $\Div$ denote the gradient and divergence operators with
respect to $\mathbf{X}$.  If there are no body forces then $\bm\tau$ must satisfy the equilibrium equation
$\Div\bm\tau=\mathbf{0}$.  For the most part we shall not be concerned with how the initial stress arises, but in Section
\ref{sec7} we shall relate $\bm\tau$ to an underlying finite deformation and $\bm\tau$ is then considered to be a
\emph{pre-stress}, which requires appropriate tractions on the boundary $\partial\mathcal{B}_r$ of $\mathcal{B}_r$.  However, if
the traction on the boundary $\partial\mathcal{B}_r$ of $\mathcal{B}_r$ vanishes pointwise then $\bm\tau$ is referred to as a
\emph{residual stress}. A residual stress is necessarily non-uniform \citep{Hoge85,Ogde03} and in general is not associated with
a deformation from a stress-free configuration.

Next, we consider the body to be subject to a finite elastic deformation from $\mathcal{B}_r$ into a new configuration
$\mathcal{B}$ with boundary $\partial\mathcal{B}$ so that the material point $\mathbf{X}$ takes up the position $\mathbf{x}$ in
$\mathcal{B}$ given by $\mathbf{x}=\bm\chi(\mathbf{X})$, where the vector function $\bm\chi$ defines the deformation for
$\mathbf{X}\in\mathcal{B}_r$.  The \emph{deformation} $\bm\chi$ is required to be a bijection and to possess appropriate
regularity properties, which we do not need to specify here.  The deformation gradient tensor, denoted $\mathbf{F}$, is defined
by $\mathbf{F}=\Grad\bm\chi$, from which are formed the left and right Cauchy--Green deformation tensors, defined by
\begin{equation}
\mathbf{B}=\mathbf{FF}^\mathrm{T},\quad \mathbf{C}=\mathbf{F}^\mathrm{T}\mathbf{F},
\end{equation}
respectively.

We denote by $\bm\sigma$ the Cauchy stress tensor in the configuration $\mathcal{B}$ and we suppose that there are no body
forces, so that the equilibrium equation $\div\bm\sigma=\mathbf{0}$ holds. We shall also make use of the nominal and the second
Piola--Kirchhoff stress tensors, denoted $\mathbf{S}$ and $\mathbf{T}$, respectively, which are related to $\bm\sigma$ and each
other by
\begin{equation}
\mathbf{S}=J\mathbf{F}^{-1}\bm\sigma=\mathbf{T}\mathbf{F}^\mathrm{T}, \quad \mathbf{T}=J\mathbf{F}^{-1}\bm\sigma\mathbf{F}^{-\mathrm{T}}
= \mathbf{SF}^{-T},\label{stressconnections}
\end{equation}
where $J=\det\mathbf{F}>0$.
The nominal stress $\mathbf{S}$ satisfies the equilibrium equation
\begin{equation}
\Div\mathbf{S}=\mathbf{0}.\label{Sequilibrium}
\end{equation}
In the absence of intrinsic couples, $\bm\sigma$, and hence $\mathbf{T}$, is symmetric, while in general $\mathbf{S}$ is not
symmetric and satisfies
\begin{equation}
\mathbf{FS}=\mathbf{S}^\mathrm{T}\mathbf{F}^\mathrm{T}.\label{symmetry}
\end{equation}

We now consider the elastic properties of the material to be characterized in terms of a strain-energy function, defined per unit
volume in $\mathcal{B}_r$, which we denote by $W$. We write
\begin{equation}
W=W(\mathbf{F},\bm\tau)\label{WFtau}
\end{equation}
to reflect the dependence not only on the deformation gradient but also on the initial stress.  By objectivity $W$ depends on
$\mathbf{F}$ only through $\mathbf{C}=\mathbf{F}^\mathrm{T}\mathbf{F}$, but it is convenient to retain the functional dependence
indicated in \eqref{WFtau}. In general the presence of the initial stress will generate anisotropy in the material response
relative to $\mathcal{B}_r$ and $\bm\tau$ has a role similar to that of a structure tensor associated with a preferred direction
in an anisotropic material. An exception to this arises if $\bm\tau$ is an isotropic stress. If $\bm\tau$ is non-uniform then the
material is necessarily inhomogeneous, but if $\bm\tau$ is independent of $\mathbf{X}$ the material is homogeneous unless its
properties depend separately on $\mathbf{X}$. In the present paper we shall consider $\bm\tau$ to be uniform and the material to
be homogeneous.

We shall not consider internal constraints such as incompressibility, in which case the nominal stress is given by
\begin{equation}
\mathbf{S}=\frac{\partial W}{\partial\mathbf{F}}(\mathbf{F},\bm\tau),\label{stress-comp}
\end{equation}
and the Cauchy and second Piola--Kirchhoff stresses can be obtained from \eqref{stressconnections}. When evaluated in
$\mathcal{B}_r$, \eqref{stress-comp} reduces to
\begin{equation}
\bm\tau=\frac{\partial W}{\partial\mathbf{F}}(\mathbf{I},\bm\tau),\label{stress-comp-ref}
\end{equation}
where $\mathbf{I}$ is the identity tensor.  Equation \eqref{stress-comp-ref} imposes a restriction on the admissible forms of
strain-energy function for an initially-stressed elastic material.

%%%%%%%%%%%%%%%%%%%%%%%%%%%%%%%%%%%%%%%%%%

\subsection{Invariant representation of the strain energy and stresses\label{sec2-1}}

%%%%%%%%%%%%%%%%%%%%%%%%%%%%%%%%%%%%%%%%%%

The strain-energy function $W$ depends on $\mathbf{C}$ and $\bm\tau$, both of which are independent of rotations in the deformed
configuration $\mathcal{B}$.  Thus, $W$ is automatically objective.  If the material possesses no intrinsic anisotropy relative
to $\mathcal{B}_r$, so that it would be isotropic relative to $\mathcal{B}_r$ in the absence of initial stress, then $W$ is an
isotropic function of $\mathbf{C}$ and $\bm\tau$, i.e.
\begin{equation}
W(\mathbf{Q}\mathbf{C}\mathbf{Q}^\mathrm{T}, \mathbf{Q}\bm\tau\mathbf{Q}^\mathrm{T})=W(\mathbf{C},\bm\tau)
\quad \mbox{for all orthogonal}\ \mathbf{Q},
\end{equation}
and it can be expressed as a function of the invariants of $\mathbf{C}$ and $\bm\tau$. We list a possible (and complete) set of
independent invariants as
\begin{align}
& I_1=\tr\mathbf{C},\quad I_2=\tfrac{1}{2}[(\tr\mathbf{C})^2-\tr(\mathbf{C}^2)],\quad I_3=\det\mathbf{C},\label{invs1}\\[1ex]
&\tr\bm\tau,\quad \tr(\bm\tau^2),\quad \tr(\bm\tau^3),\label{invs2}\\[1ex]
& I_6=\tr(\bm\tau\mathbf{C}),\quad I_7= \tr(\bm\tau\mathbf{C}^2),\quad I_8=\tr(\bm\tau^2\mathbf{C}),\quad I_9=\tr(\bm\tau^2\mathbf{C}^2), \label{invs3}
\end{align}
where we have used the standard notation $I_1,I_2,I_3$ for the principal invariants of $\mathbf{C}$ and followed the notation
$I_6,\ldots,I_9$ adopted by \cite{Sham11} for the combined invariants of $\mathbf{C}$ and $\bm\tau$.  In the reference
configuration $\mathcal{B}_r$ these reduce to
\begin{equation}
I_1=I_2=3,\quad I_3=1,\quad I_6=I_7=\tr\bm\tau,\quad I_8=I_9=\tr(\bm\tau^2).\label{Isref}
\end{equation}
For full discussion of invariants of tensors we refer to \citet{Spen71} and \citet{Zhen94}.  Here there are 10 independent
invariants of $\mathbf{C}$ and $\bm\tau$ in general, a number that may be reduced in a two-dimensional specialization or for
specific simple deformations and/or initial stresses. We have not attributed notations to the invariants \eqref{invs2} since they
are independent of the deformation and do not contribute explicitly to expressions for the stresses.  However, $W$ may depend on
\eqref{invs2} implicitly, but we do not list them in the functional dependence and we write $W=W(I_1,I_2,I_3,I_6, I_7,I_8,I_9)$,
retaining the notation $W$, which is used severally for the $(\mathbf{F},\bm\tau)$, $(\mathbf{C},\bm\tau)$ and ($I_1$, $I_2$,
$I_3$, $I_6,I_7,I_8,I_9$) arguments.

From \eqref{stress-comp} the nominal stress tensor may be expanded as
\begin{equation}
\mathbf{S}=\frac{\partial W}{\partial \mathbf{F}}=\sum_{i\,\in\,\mathcal{I}}W_i\frac{\partial I_i}{\partial \mathbf{F}},
\end{equation}
where we have used the shorthand notation $W_i=\partial W/\partial I_i,\, i\in\mathcal{I}$, and $\mathcal{I}$ is the index set
$\{1,2,3,6,7,8,9\}$.  We emphasize that although their derivatives with respect to $\mathbf{F}$ vanish the invariants
\eqref{invs2} are included implicitly in the functional dependence of $W$ in general.  The required expressions for $\partial
I_i/\partial \mathbf{F}$ were given in Appendix A of \citep{Sham11} and are not repeated here explicitly but used implicitly. The
resulting expression for the Cauchy stress $\bm\sigma$ is obtained from \eqref{stressconnections} in the form
\begin{eqnarray}
J\bm\sigma &=& 2 W_1 \mathbf{B} + 2 W_2 (I_1 \mathbf{B} - \mathbf{B}^2) + 2I_3W_3 \mathbf{I} + 2 W_6 \bm\Sigma\notag
\\[1ex]
 &+& 2 W_7 (\bm\Sigma\mathbf{B} + \mathbf{B}\bm\Sigma)
 + 2W_8 \bm\Sigma \mathbf{B}^{-1} \bm\Sigma
 + 2 W_9(\bm\Sigma\mathbf{B}^{-1} \bm\Sigma\mathbf{B}
   + \mathbf{B} \bm\Sigma \mathbf{B}^{-1} \bm\Sigma),\label{comp-sigma}
\end{eqnarray}
wherein we have introduced the notation $\bm\Sigma=\mathbf{F}\bm\tau\mathbf{F}^\mathrm{T}$ for the push forward of $\bm\tau$ from
$\mathcal{B}_r$ to $\mathcal{B}$, and $\mathbf{B}=\mathbf{FF}^\mathrm{T}$ is the left Cauchy--Green tensor.

When \eqref{comp-sigma} is evaluated in the reference configuration it reduces to
\begin{equation}
\bm\tau = 2(W_1 + 2W_2 +W_3)\mathbf{I} + 2(W_6 + 2 W_7)\bm\tau + 2(W_8 + 2 W_9)\bm\tau^{2},\label{sigma-in-ref-comp}
\end{equation}
where $W_i,\,i\in\mathcal{I}$, are evaluated for the invariants given by
\eqref{Isref}. Equation \eqref{sigma-in-ref-comp} is the specialization of \eqref{stress-comp-ref} for the invariant form of $W$.
As in \cite{Sham11} we deduce that
\begin{equation}
W_1 + 2 W_2 +W_3 = 0, \quad 2(W_6 + 2 W_7) = 1, \quad W_8 + 2 W_9 = 0\label{ref-conds-comp}
\end{equation}
in $\mathcal{B}_r$.

%%%%%%%%%%%%%%%%%%%%%%%%%%%%%%%%%%%%%%%%%%

\section{Incremental motions\label{sec3}}

%%%%%%%%%%%%%%%%%%%%%%%%%%%%%%%%%%%%%%%%%%

We now consider the static finite deformation $\mathbf{x}=\bm\chi(\mathbf{X})$ that defines the equilibrium configuration
$\mathcal{B}$ to be followed by a superimposed incremental motion $\mathbf{\dot{x}}(\mathbf{X},t)$, where $t$ is time. A
superposed dot signifies an incremental quantity and the resulting incremental equations are linearized in the increments, which
are considered appropriately `small'. Thus, $\mathbf{\dot{x}}$ represents a small displacement from $\mathbf{x}$. We shall also
write the displacement in Eulerian form as $\mathbf{u}=\mathbf{u}(\mathbf{x},t)$, noting that
$\mathbf{\dot{x}}(\mathbf{X},t)=\mathbf{u}(\bm\chi(\mathbf{X}),t)$. The corresponding increments in the deformation gradient
$\mathbf{F}$ and $J=\det\mathbf{F}$ are given by the standard formulas
\begin{equation}
\mathbf{\dot{F}}=\mathbf{LF},\quad \dot{J}=J\tr\mathbf{L},\label{Fdot}
\end{equation}
where $\mathbf{L}=\grad\mathbf{u}$ is the displacement gradient.

The (linearized) incremental nominal stress is written
\begin{equation}
\mathbf{\dot{S}}= \bm{\mathcal{A}}\mathbf{\dot{F}}, \label{incrementalS}
\end{equation}
where
\begin{equation}
\bm{\mathcal{A}}
= \frac{\partial^2W}{\partial\mathbf{F}\partial\mathbf{F}},
\quad
\mathcal{A}_{\alpha i\beta j}=\frac{\partial^2W}{\partial F_{i\alpha}\partial F_{j\beta}},\label{mathcalA-def}
\end{equation}
is the fourth-order \emph{elasticity tensor} and, in component form, $\bm{\mathcal{A}}\mathbf{\dot{F}}
\equiv \mathcal{A}_{\alpha i\beta j}\dot{F}_{j\beta}$ defines the product used in \eqref{incrementalS}.
The usual summation convention for repeated indices is adopted here and henceforth. For full discussion of the theory of incremental
deformations and motions superimposed on a finite deformation we refer to \citet{Ogde84,Ogde07}, for example.

By taking the increments of the connections $J\bm\sigma=\mathbf{FS}$ and $\mathbf{S}=\mathbf{T}\mathbf{F}^\mathrm{T}$ from
\eqref{stressconnections} we obtain, after a little rearrangement,
\begin{eqnarray}
\mathbf{\dot{S}}_0&\equiv&
J^{-1}\mathbf{F}\mathbf{\dot{S}}=\mathbf{\dot{\bm\sigma}}+(\tr\mathbf{L})\bm\sigma-\mathbf{L}\bm\sigma,\label{inrementalSsigma}\\[1ex]
\mathbf{\dot{T}}_0&\equiv&J^{-1}\mathbf{F}\mathbf{\dot{T}}\mathbf{F}^\mathrm{T}=\mathbf{\dot{S}}_0-\bm\sigma\mathbf{L}^\mathrm{T},\label{inrementalTS}
\end{eqnarray}
wherein the notations $\mathbf{\dot{S}}_0$ and $\mathbf{\dot{T}}_0$ are defined.  These are the updated forms of $\mathbf{\dot{S}}$ and
$\mathbf{\dot{T}}$, respectively, referred to deformed configuration, and otherwise know as their `push forward' forms.  The corresponding push forward $\bm{\mathcal{A}}_0$ of the elasticity tensor is
such that $\mathbf{\dot{S}}_0=\bm{\mathcal{A}}_0\mathbf{L}$.  It then follows from the symmetry of $\bm\sigma$ and its increment
that $\mathbf{\dot{S}}_0+\mathbf{L}\bm\sigma$ is symmetric, and hence
\begin{equation}
\bm{\mathcal{A}}_0\mathbf{L}+\mathbf{L}\bm\sigma=(\bm{\mathcal{A}}_0\mathbf{L})^\mathrm{T}+\bm\sigma\mathbf{L}^\mathrm{T}.\label{symmetry3}
\end{equation}
In component form $\bm{\mathcal{A}}_0$ is related to $\bm{\mathcal{A}}$ via
\begin{equation}
J\mathcal{A}_{0piqj}=F_{p\alpha}F_{q\beta}\mathcal{A}_{\alpha i\beta j}.\label{AA0connect}
\end{equation}

Note that as well as possessing the major symmetry $\mathcal{A}_{0piqj}=\mathcal{A}_{0qjpi}$, which follows from
\eqref{mathcalA-def} and \eqref{AA0connect}, $\bm{\mathcal{A}}_0$ has the property
\begin{eqnarray} \label{property}
\mathcal{A}_{0piqj}+\delta_{jp}\sigma_{iq}=\mathcal{A}_{0ipqj}+\delta_{ij}\sigma_{pq},
\end{eqnarray}
which can be deduced from \eqref{symmetry3}.

We assume that there are no body forces. Then, the incremental motion is governed by the equation
\begin{eqnarray}
\Div\mathbf{\dot{S}}=\rho_r\mathbf{x}_{,tt},\label{motion1}
\end{eqnarray}
where $\rho_r$ is the mass density in $\mathcal{B}_r$ and a subscript $t$ following a comma signifies
the material time derivative, i.e. the time derivative at fixed $\mathbf{X}$, so that
$\mathbf{x}_{,t}=\mathbf{u}_{,t}$ is the particle velocity and $\mathbf{x}_{,tt}=\mathbf{u}_{,tt}$
the acceleration.

Equation \eqref{motion1} may be updated (i.e. pushed forward) to the configuration $\mathcal{B}$ by writing it in terms of
$\mathbf{\dot{S}}_0$ and $\mathbf{u}$, which yields
\begin{equation}
\div\mathbf{\dot{S}}_0=\rho\mathbf{u}_{,tt},
\end{equation}
where $\rho=\rho_rJ^{-1}$ is the mass density in $\mathcal{B}$, or, in (Cartesian) component form,
\begin{equation}
(\mathcal{A}_{0piqj}u_{j,q})_{,p}=\rho u_{i,tt}.\label{motioncomp}
\end{equation}

Henceforth in this paper we assume that the initial stress $\bm\tau$, the underlying deformation $\mathbf{F}$ and the material
properties are homogeneous, so that $\bm{\mathcal{A}}$ and $\bm{\mathcal{A}}_0$ are independent of $\mathbf{X}$. The equation of
motion \eqref{motioncomp} then becomes
\begin{eqnarray}
\mathcal{A}_{0piqj}u_{j,pq}=\rho u_{i,tt}.\label{motionhomogeneous}
\end{eqnarray}
This will be used in Section \ref{sec6} in discussion of the propagation of
homogeneous plane waves, but before proceeding to that analysis we obtain explicit expressions for the dependence of the
components of the elasticity tensor and of various elastic moduli on the initial stress based on the invariants of the right
Cauchy--Green deformation tensor $\mathbf{C}$ and the initial stress tensor $\bm\tau$ discussed in Section \ref{sec2-1}.

%%%%%%%%%%%%%%%%%%%%%%%%%%%%%%%%%%%%%%%%%%

\section{Expressions for the elasticity tensor\label{sec4}}

%%%%%%%%%%%%%%%%%%%%%%%%%%%%%%%%%%%%%%%%%%

The elasticity tensor $\bm{\mathcal{A}}$ in \eqref{mathcalA-def} may be expanded in terms of invariants as
\begin{eqnarray}
\bm{\mathcal{A}}
=\sum_{i\,\in\,\mathcal{I}}W_i\frac{\partial^2I_i}{\partial\mathbf{F}\partial\mathbf{F}}
+\sum_{i,j\,\in\,\mathcal{I}}W_{ij}\frac{\partial I_i}{\partial\mathbf{F}}\otimes\frac{\partial I_j}{\partial\mathbf{F}},
\end{eqnarray}
where  $W_{ij}
=\partial^2 W/\partial I_i\partial I_j,\, i,j\in\mathcal{I}$, and $\mathcal{I}$ is again the index set
$\{1,2,3,6,7,8,9\}$.

The detailed (lengthy) expressions for the components of $\bm{\mathcal{A}}_0$ were given by \cite{Sham11} for a general
deformed configuration based on expressions for the second derivatives of the invariants, which were given in
Appendix A of the latter paper.  Here we require only their specialization to the (undeformed) reference configuration ($\mathcal{B}\rightarrow\mathcal{B}_r$),
which, following \cite{Sham11}, yields
\begin{multline} \label{A0-comp-components-ref}
\mathcal{A}_{0piqj}
=
 \alpha_1 (\delta_{ij}\delta_{pq}+\delta_{iq}\delta_{jp})
+
\alpha_2 \delta_{ip}\delta_{jq}+\delta_{ij}\tau_{pq}+\beta_1(\delta_{ij}\tau_{pq}+\delta_{pq}\tau_{ij}
+\delta_{iq}\tau_{jp}+\delta_{jp}\tau_{iq})
\\[1ex]
+
\beta_2(\delta_{ip}\tau_{jq} +\delta_{jq}\tau_{ip})
+
\beta_3\tau_{ip}\tau_{jq}+\gamma_1(\delta_{ij}\tau_{pk}\tau_{kq} + \delta_{pq}\tau_{ik}\tau_{kj}
+ \delta_{iq}\tau_{jk}\tau_{kp} + \delta_{jp}\tau_{ik}\tau_{kq})
\\[1ex]
+
\gamma_2(\delta_{ip}\tau_{jk}\tau_{kq} + \delta_{jq}\tau_{ik}\tau_{kp})+\gamma_3(\tau_{ip}\tau_{jk}\tau_{kq}
+\tau_{jq}\tau_{ik}\tau_{kp})
+
\gamma_4\tau_{ik}\tau_{kp}\tau_{jl}\tau_{lq},
\end{multline}
where the $\alpha$'s, $\beta$'s, and $\gamma$'s are defined by
\begin{eqnarray}
 \alpha_1 &=& 2(W_1+W_2),\quad
\alpha_2 = 4(W_{11}+4W_{12}+4W_{22}+2W_{13}+4W_{23}+W_{33} - W_1-W_2),
 \notag \\[1ex]
\beta_1 &=& 2W_7,\quad
\beta_2 = 4(W_{16}+2W_{17}+2W_{26}+4W_{27}+W_{36}+2W_{37}),\quad \gamma_1= 2W_9,
\notag \\[1ex]
\beta_3& =& 4(W_{66}+4W_{67}+4W_{77}), \quad
\gamma_2 = 4(W_{18} + 2W_{19} + 2W_{28}+4W_{29}+W_{38} + 2W_{39}),\notag\\[1ex]
\gamma_3&=&4(W_{68}+2W_{69}+2W_{78}+4W_{79}),
\quad
\gamma_4 = 4(W_{88} + 4W_{89} + 4W_{99}),
\label{alpha--epsilon}
\end{eqnarray}
all derivatives $W_i$ and $W_{ij}$ being evaluated in the reference configuration and use having been made of the connections
\eqref{ref-conds-comp}. Note that in general the expressions \eqref{alpha--epsilon} may depend on the invariants $\tr\bm\tau$,
$\tr(\bm\tau^2)$ and $\tr(\bm\tau^3)$.  There are nine ($\bm\tau$-dependent) material parameters in the above, just as there are nine constants for an
orthotropic linearly elastic material (see, for example, \citealp{Ting96}), but additionally here the components $\tau_{ij}$ of
$\bm\tau$ are present separately.

Note that when referred to axes that coincide with the principal axes of $\bm\tau$, the only
non-zero components of \eqref{A0-comp-components-ref} are given by
\begin{align}
\label{elasticity-tensor-principal-tau}
&
\mathcal{A}_{0iiii}
= 2\alpha_1+\alpha_2+(1+4\beta_1+2\beta_2)\tau_i+\beta_3\tau_i^2+2(2\gamma_1+\gamma_2)\tau_i^2+2\gamma_3\tau_i^3+\gamma_4\tau_i^4,
\notag \\[1ex]
&
\mathcal{A}_{0iijj}
=\alpha_2+\beta_2(\tau_i+\tau_j)+\beta_3\tau_i\tau_j+\gamma_2(\tau_i^2+\tau_j^2)+\gamma_3(\tau_i+\tau_j)\tau_i\tau_j+\gamma_4\tau_i^2\tau_j^2,
\notag \\[1ex]
&
\mathcal{A}_{0ijij}
=\alpha_1+\tau_i+\beta_1(\tau_i+\tau_j)+\gamma_1(\tau_i^2+\tau_j^2)
=\mathcal{A}_{0ijji}+\tau_i,
\end{align}
where there are no sums on repeated indices, $i\neq j$, and $\tau_i$, $i=1,2,3$, are the principal values of $\bm\tau$ (in
general, there are 15 non-zero components of $\bm{\mathcal{A}_0}$ in total, dependent on nine material parameters and three
principal stresses).  In the linear specialization of the above only the parameters $\alpha_1,\alpha_2,\beta_1,\beta_2$ are
retained along with $\tau_1,\tau_2,\tau_3$, although we should strictly expand $\alpha_1$ and $\alpha_2$ as linear functions of
$\bm\tau$. Then there are 9 constants involved, specifically $\alpha_1(\mathbf{0})$, $\alpha_2(\mathbf{0})$, $\beta_1(\mathbf{0})$, $\beta_2(\mathbf{0})$, $\alpha_{1,i}(\mathbf{0})$,
$\alpha_{2,i}(\mathbf{0}),\,i=1,2,3,$ and the three principal initial stresses $\tau_1$, $\tau_2$, $\tau_3$, where $\mathbf{0}=(0,0,0)$ is the value of $(\tau_1, \tau_2, \tau_3)$ for zero initial stress and $_{,i}$ signifies differentiation with respect to $\tau_i,\,i=1,2,3$. Note that since the coefficients in \eqref{elasticity-tensor-principal-tau} are symmetric functions of $(\tau_1, \tau_2, \tau_3)$ the constants $\alpha_{1,i}(\mathbf{0})$ and $\alpha_{2,i}(\mathbf{0})$ are independent of $i$.

We now specialize the above to consider three specific forms of $\bm\tau$, corresponding to isotropic initial stress, uniaxial
initial stress and planar shear initial stress.

%%%%%%%%%%%%%%%%%%%%%%%%%%%%%%%%%%%%%%%%%%

\subsection{Isotropic initial stress}

%%%%%%%%%%%%%%%%%%%%%%%%%%%%%%%%%%%%%%%%%%

Suppose that $\bm\tau$ is isotropic and write $\bm\tau=\tau\mathbf{I}$, where $\mathbf{I}$ is again the identity tensor and
$\tau>0 \,(<0)$ corresponds to hydrostatic tension (pressure).  Then equation \eqref{A0-comp-components-ref} reduces to the
compact form
\begin{equation}
\mathcal{A}_{0piqj}=\tau\delta_{ij}\delta_{pq}+\alpha(\tau)(\delta_{ij}\delta_{pq}+\delta_{iq}\delta_{pj})
+\beta(\tau)\delta_{pi}\delta_{qj},\label{isocalA}
\end{equation}
where the notations $\alpha(\tau)$ and $\beta(\tau)$ are defined by
\begin{eqnarray}
\alpha(\tau)&=&\alpha_1(\tau)+2\tau\beta_1(\tau)+2\tau^2\gamma_1(\tau),\\[1ex]
\beta(\tau)&=&\alpha_2(\tau)+2\tau\beta_2(\tau)+\tau^2[\beta_3(\tau)+2\gamma_2(\tau)]+2\tau^3\gamma_3(\tau)+\tau^4\gamma_4(\tau),
\end{eqnarray}
and we note that, by virtue of the specializations $\tr\bm\tau=3\tau,\tr(\bm\tau^2)=3\tau^2,\tr(\bm\tau^3)=3\tau^3$, the
coefficients $\alpha_1,...,\gamma_4$ are now (in general) functions of the single parameter $\tau$, which is indicated above by
inclusion of the argument $\tau$.

%%%%%%%%%%%%%%%%%%%%%%%%%%%%%%%%%%%%%%%%%%

\subsection{Uniaxial initial stress}

%%%%%%%%%%%%%%%%%%%%%%%%%%%%%%%%%%%%%%%%%%

Here we take $\bm\tau=\tau \mathbf{a}\otimes\mathbf{a}$, where $\mathbf{a}$ is a fixed unit vector (the direction
of the uniaxial stress, which is a tension for $\tau >0$ and a compressive stress for $\tau<0$).  In this case the
components of $\bm{\mathcal{A}_0}$ may be expressed in the form
\begin{multline}
\mathcal{A}_{0piqj}=\alpha_1(\delta_{ij}\delta_{pq}+\delta_{iq}\delta_{pj})+\alpha_2\delta_{pi}\delta_{qj}
+\tau(\beta_1+\tau\gamma_1)(\delta_{ij}a_pa_q+\delta_{pq}a_ia_j+\delta_{iq}a_pa_j+\delta_{pj}a_ia_q)\\[1ex]
+\tau\delta_{ij}a_pa_q +\tau(\beta_2+\tau\gamma_2)(\delta_{pi}a_qa_j+\delta_{qj}a_pa_i)+\tau^2(\beta_3+2\tau\gamma_3
+\tau^2\gamma_4)a_pa_ia_qa_j.
\end{multline}
Again the coefficients $\alpha_1,...,\gamma_4$ depend on $\tau$ in general but for the sake of brevity this has been left
implicit here. Note that, in addition to $\tau$, there are five separate (combinations of) parameters, namely $\alpha_1$,
$\alpha_2$, $\beta_1+\tau\gamma_1$, $\beta_2+\tau\gamma_2$, $\beta_3+\tau\gamma_3+\tau^2\gamma_4$ and we recall that in classical
transversely isotropic linear elasticity there are five material constants (see, for example, \citealp{Ting96}).

Without loss of generality we may take $\mathbf{a}$ to coincide with the axis $\mathbf{e}_1$.  Then the components of
$\bm{\mathcal{A}_0}$ are listed as
\begin{align}
& \mathcal{A}_{01111} = 2\alpha_1 + \alpha_2 + (1 + 4\beta_1 + 2\beta_2)\tau + (4\gamma_1 + 2\gamma_2
+ \beta_3)\tau^2 + 2\gamma_3 \tau^3 + \gamma_4 \tau^4,
\\[1ex]
& \mathcal{A}_{0iiii} = 2\alpha_1+\alpha_2,
\quad
\mathcal{A}_{0iijj}=\alpha_2,
\quad
\mathcal{A}_{011ii} = \alpha_2 + \beta_2 \tau + \gamma_2 \tau^2,
\quad i,j\in\{2,3\},i\neq j,\label{symmetry23}
\\[1ex]
& \mathcal{A}_{0i1i1} = \mathcal{A}_{01ii1}=\mathcal{A}_{0i11i}=\alpha_1 + \beta_1\tau + \gamma_1\tau^2,\quad i\in\{2,3\},
\\[1ex]
& \mathcal{A}_{01i1i} = \alpha_1+  (1+\beta_1) \tau + \gamma_1\tau^2,\quad
\mathcal{A}_{0ijij}=\mathcal{A}_{0ijji}=\alpha_1,\quad i,j\in\{2,3\}, i\neq j,
\label{A2323}
\end{align}
for later reference.
Note that in the linear specialization there remain only seven independent constants, namely $\alpha_1(0)$,
$\alpha_2(0)$, $\beta_1(0)$, $\beta_2(0)$, $\alpha_1'(0)$, $\alpha_2'(0)$, with argument $\tau=0$, and $\tau$, where the prime indicates differentiation with respect to $\tau$.

%%%%%%%%%%%%%%%%%%%%%%%%%%%%%%%%%%%%%%%%%%

\subsection{Planar shear initial stress\label{sec4-3}}

%%%%%%%%%%%%%%%%%%%%%%%%%%%%%%%%%%%%%%%%%%

Consider planar shear stress in the $(x_1,x_2)$ plane of the form $\bm\tau=\tau(\mathbf{e}_1\otimes\mathbf{e}_2
+\mathbf{e}_2\otimes\mathbf{e}_1)$.  Then the only non-zero components of $\bm{\mathcal{A}}_0$ are written as
\begin{align}
& \mathcal{A}_{01111} = \mathcal{A}_{02222}=2\alpha_1+\alpha_2+2(2\gamma_1+\gamma_2)\tau^2+\gamma_4\tau^4,
\qquad
\mathcal{A}_{03333}=2\alpha_1+\alpha_2,
\\[1ex]
& \mathcal{A}_{01122} = \alpha_2+2\gamma_2\tau^2+\gamma_4\tau^4, \quad \mathcal{A}_{01133}=\mathcal{A}_{02233}=\alpha_2,
\\[1ex]
&\mathcal{A}_{0ijij} = \mathcal{A}_{0ijji}=\alpha_1+(\beta_3+2\gamma_1)\tau^2,\quad i,j\in\{1,2\},i\neq j,
\\[1ex]
& \mathcal{A}_{0i3i3} = \mathcal{A}_{03i3i}=\mathcal{A}_{03ii3}=\alpha_1+\gamma_1\tau^2,\quad i\in\{1,2\},
\\[1ex]
& \mathcal{A}_{0iiij} = (2\beta_1+\beta_2)\tau+\gamma_3\tau^3,\quad \mathcal{A}_{0jiii}=\mathcal{A}_{0iiij}
+\tau,\quad i,j\in\{1,2\},i\neq j.
\end{align}
In the linear specialization there are now five constants, $\alpha_1(0)$, $\alpha_2(0)$, $2\beta_1(0) + \beta_2(0)$,
$\alpha_1'(0)$, $\alpha_2'(0)$, in addition to $\tau$.

%%%%%%%%%%%%%%%%%%%%%%%%%%%%%%%%%%%%%%%%%%

\section{Dependence of elastic moduli on initial stress\label{sec5}}

%%%%%%%%%%%%%%%%%%%%%%%%%%%%%%%%%%%%%%%%%%

\subsection{Isotropic initial stress}

%%%%%%%%%%%%%%%%%%%%%%%%%%%%%%%%%%%%%%%%%%

In the connection \eqref{inrementalSsigma} we now specialize the Cauchy stress $\bm\sigma$ to the initial stress $\bm\tau$, so that
\begin{equation}
\mathbf{\dot{\bm\sigma}}=\mathbf{\dot{S}}_0+\mathbf{L}\bm\tau-(\tr\mathbf{L})\bm\tau.\label{Sdotsigmadotconnection}
\end{equation}
It follows on use of \eqref{isocalA} that, for an isotropic initial stress $\bm\tau = \tau \mathbf{I}$,
\begin{equation}
\mathbf{\dot{S}}_{0}=\bm{\mathcal{A}}_{0}\mathbf{L}=\tau \mathbf{L}^\mathrm{T}+\alpha(\tau)(\mathbf{L}
+\mathbf{L}^\mathrm{T})+\beta(\tau)(\tr\mathbf{L})\mathbf{I},\label{dotSiso}
\end{equation}
and hence that
\begin{equation}
\mathbf{\dot{\bm\sigma}} =  [\beta(\tau)-\tau](\tr\mathbf{L})\mathbf{I}
 + [\alpha(\tau)+\tau](\mathbf{L}+\mathbf{L}^\mathrm{T}).
\end{equation}

Since the initial stress is purely isotropic we can therefore identify the \emph{stress-dependent Lam\'e moduli}, which we denote
as  $\lambda(\tau)$ and $\mu(\tau)$, as
\begin{align}
& \lambda(\tau)=\beta(\tau)-\tau=\alpha_2(\tau)+2\tau\beta_2(\tau)+\tau^2[\beta_3(\tau)+2\gamma_2(\tau)]
+2\tau^3\gamma_3(\tau)+\tau^4\gamma_4(\tau)-\tau,
\label{lambdaiso}
\\[1ex]
& \mu(\tau)=\alpha(\tau)+\tau=\alpha_1(\tau)+2\tau\beta_1(\tau)+2\tau^2\gamma_1(\tau)+\tau.
\label{muiso}
\end{align}

For incremental simple shear we may, without loss of generality, restrict attention to the $(x_1,x_2)$ plane. If the shear is in
the $x_1$ direction with amount of shear $L_{ij}=L_{12}\delta_{1i}\delta_{2j}$ then the corresponding incremental nominal and
Cauchy stress components are equal and given by $\dot{S}_{021}=\dot{\sigma}_{12}=\mathcal{A}_{02121}L_{12}=\mu(\tau)L_{12}$.
Similarly, for incremental simple shear in the $x_2$ direction with $L_{ij}=L_{21}\delta_{2i}\delta_{1j}$ we have
$\dot{S}_{012}=\dot{\sigma}_{12}=\mathcal{A}_{01212}L_{21}=\mu(\tau)L_{21}$, and from \eqref{isocalA} and \eqref{muiso},
$\mathcal{A}_{02121}=\mathcal{A}_{01212}=\mu(\tau)$.

For incremental pure dilatation with $L_{11}=L_{22}=L_{33}=(\tr\mathbf{L})/3$  we obtain
\begin{equation}
\tr\mathbf{\dot{\bm\sigma}}=[3\lambda(\tau)+2\mu(\tau)]\tr\mathbf{L},
\end{equation}
and this enables us to define the \emph{stress-dependent bulk modulus} $\kappa(\tau)$, analogously to the classical formula, as
\begin{equation}
\kappa(\tau)=\lambda(\tau)+\frac{2}{3}\mu(\tau).\label{kappaiso}
\end{equation}

Note that from \eqref{Sdotsigmadotconnection} we obtain $\tr\mathbf{\dot{S}}_0=\tr\mathbf{\dot{\bm\sigma}}+2\tau\tr\mathbf{L}$,
but use of $\tr\mathbf{\dot{S}}_0$ instead of $\tr\mathbf{\dot{\bm\sigma}}$ does not give the correct form of the bulk modulus.  Also,
by using \eqref{inrementalTS} with $\bm\sigma=\bm\tau=\tau\mathbf{I}$ and \eqref{dotSiso} we obtain
\begin{equation}
\mathbf{\dot{T}}_0=\alpha(\tau)(\mathbf{L}+\mathbf{L}^\mathrm{T})+\beta(\tau)(\tr\mathbf{L})\mathbf{I}.\label{Tdotiso}
\end{equation}
Thus, it is clear that because of the dependence on the initial stress, different choices of stress measure lead to
different possible definitions of the stress-dependent elastic moduli.

The correct definitions for the stress dependent Lam\'e moduli are \eqref{lambdaiso} and \eqref{muiso}. If the second
Piola--Kirchhoff stress is used instead, then $\mu(\tau)$ and $\lambda(\tau)$ would be replaced by $\alpha(\tau)$ and
$\beta(\tau)$, respectively. This was effectively what was done in the paper by \cite{Tang67}, although he worked in terms of
Young's modulus and Poisson's ratio. This identification of the Lam\'e moduli leads to erroneous results for the speeds of
homogeneous plane waves, as we shall show in Section \ref{sec6}.

If there is no initial stress (reduction to the classical case) we denote the classical moduli by $\lambda_0,
\mu_0,\kappa_0$, so that $\lambda_0=\lambda(0) = \alpha_2(0)$, $\mu_0 = \mu(0) = \alpha_1(0)$, and $\kappa_0=\kappa(0)$. If the
initial stress is small in magnitude then we may linearize the expressions \eqref{lambdaiso}, \eqref{muiso} and \eqref{kappaiso}
to obtain
\begin{align}
& \lambda(\tau) \simeq  \lambda_0+[\alpha_2'(0)+2\beta_2(0)-1]\tau,\\[1ex]
& \mu(\tau) \simeq  \mu_0+[\alpha_1'(0)+2\beta_1(0)+1]\tau,\\[1ex]
& \kappa(\tau) \simeq \kappa_0+[2\alpha_1'(0)+3\alpha_2'(0)+4\beta_1(0)+6\beta_2(0)-1]\tau/3.
\end{align}

%%%%%%%%%%%%%%%%%%%%%%%%%%%%%%%%%%%%%%%%%%

\subsection{Uniaxial initial stress}

%%%%%%%%%%%%%%%%%%%%%%%%%%%%%%%%%%%%%%%%%%

When the initial stress is uniaxial the subsequent incremental response is transversely isotropic in nature. Then it is
appropriate to work in terms of Young's moduli and Poisson's ratios. In order to determine these we need to examine both triaxial
incremental deformations without shear and separate incremental shear deformations. Consider first the normal components of
$\mathbf{L}$, written $L_{11},L_{22},L_{33}$, and take $\mathbf{e}_1$ to be the direction of the uniaxial initial stress. Then,
bearing in mind the symmetry in \eqref{symmetry23}, the corresponding incremental nominal stresses are
\begin{eqnarray}
\dot{S}_{011}&=&\mathcal{A}_{01111}L_{11}+\mathcal{A}_{01122}(L_{22}+L_{33}),\\[1ex]
\dot{S}_{022}&=&\mathcal{A}_{01122}L_{11}+\mathcal{A}_{02222}L_{22}+\mathcal{A}_{02233}L_{33},\\[1ex]
\dot{S}_{033}&=&\mathcal{A}_{01122}L_{11}+\mathcal{A}_{02233}L_{22}+\mathcal{A}_{02222}L_{33}.
\end{eqnarray}
The (incremental) Poisson's ratio $\nu_{12}$ (and hence $\nu_{13}$ by symmetry) is obtained by setting $\dot{S}_{022}=\dot{S}_{033}=0$ and
using the resulting symmetry $L_{33}=L_{22}$ and $L_{22}=-\nu_{12}L_{11}$ to obtain
\begin{equation}
\nu_{12}=\mathcal{A}_{01122}/(\mathcal{A}_{02222}+\mathcal{A}_{02233}).
\end{equation}
Then,
\begin{equation}
\dot{S}_{011}=(\mathcal{A}_{01111}-2\nu_{12}\mathcal{A}_{01122})L_{11}
\end{equation}
and the (incremental) Young's modulus $E_1$ can be read off as
\begin{equation}
E_1=\mathcal{A}_{01111}-2\nu_{12}\mathcal{A}_{01122}=\mathcal{A}_{01111}-2\mathcal{A}_{01122}^2/(\mathcal{A}_{02222}+\mathcal{A}_{02233}).
\end{equation}

To obtain $\nu_{21}=\nu_{31}$, $\nu_{23}=\nu_{32}$ and $E_2=E_3$, on the other hand, we set $\dot{S}_{011}=\dot{S}_{033}=0$, with
$L_{11}=-\nu_{21}L_{22},L_{33}=-\nu_{23}L_{22}$, and hence
\begin{equation}
\nu_{21}\mathcal{A}_{01111}+\nu_{23}\mathcal{A}_{01122}=\mathcal{A}_{01122},\quad
\nu_{21}\mathcal{A}_{01122}+\nu_{23}\mathcal{A}_{02222}=\mathcal{A}_{02233}
\end{equation}
and
\begin{equation}
\dot{S}_{022}=(\mathcal{A}_{02222}-\nu_{21}\mathcal{A}_{01122}-\nu_{23}\mathcal{A}_{02233})L_{22}=E_2L_{22}.
\end{equation}
These yield
\begin{equation}
\nu_{21}=\frac{\mathcal{A}_{01122}(\mathcal{A}_{02222}-\mathcal{A}_{02233})}{\mathcal{A}_{01111}\mathcal{A}_{02222}-\mathcal{A}_{01122}^2},\quad
\nu_{23}=\frac{\mathcal{A}_{01111}\mathcal{A}_{02233}-\mathcal{A}_{01122}^2}{\mathcal{A}_{01111}\mathcal{A}_{02222}-\mathcal{A}_{01122}^2}
\end{equation}
and
\begin{equation}
E_2=(\mathcal{A}_{02222}-\mathcal{A}_{02233}) \frac{
\mathcal{A}_{01111}(\mathcal{A}_{02222}+\mathcal{A}_{02233})-2\mathcal{A}_{01122}^2}{\mathcal{A}_{01111}\mathcal{A}_{02222}-\mathcal{A}_{01122}^2}.
\end{equation}

The connection
\begin{equation}
E_2/\nu_{21}=E_1/\nu_{12}\label{nuEconnection}
\end{equation}
then follows, as in the classical linear theory. We note, however, that if the increments of the Cauchy stress were used in the
definitions of the stress-dependent Poisson's ratios and Young's moduli instead of the nominal stress (which is entirely legitimate), then this would not follow.
There are now four independent material parameters: $\nu_{12}=\nu_{13}$, $\nu_{21}=\nu_{31}$, $\nu_{23}=\nu_{32}$ and $E_1$, for
example, with $E_2=E_3$ given by \eqref{nuEconnection}.

For the incremental shear response in the plane of symmetry, we have
\begin{equation}
\dot{\sigma}_{23}=\dot{S}_{023}=\mathcal{A}_{02323}L_{32}+\mathcal{A}_{02332}L_{23}=\alpha_1(L_{23}+L_{32}),
\end{equation}
according to \eqref{A2323}$_2$, and hence $\alpha_1$ is the shear modulus in the plane of symmetry.  In fact, it is
straightforward to show that it can be expressed in terms of the other parameters as
\begin{equation}
\alpha_1=E_2/2(1+\nu_{23}),
\end{equation}
similarly to the situation in the classical theory.

However, when it comes to shear in a plane that contains the axis
$\mathbf{e}_1$ the situation differs from the classical one because of the influence of $\tau$.
We have
\begin{equation}
\dot{\sigma}_{12}=\dot{S}_{012}=\dot{S}_{021}+\tau
L_{21}=[\alpha_1+\tau(\beta_1+\tau\gamma_1)+\tau]L_{21}+[\alpha_1+\tau(\beta_1+\tau\gamma_1)]L_{12}.
\end{equation}
For shear in the $x_1$ direction with incremental simple shear $L_{ij}=L_{12}\delta_{1i}\delta_{2j}$ we obtain
\begin{equation}
\dot{\sigma}_{12}=\dot{S}_{021}=[\alpha_1+\tau(\beta_1+\tau\gamma_1)]L_{12},
\end{equation}
while for shear transverse to the $x_1$ direction with incremental  simple shear $L_{ij}=L_{21}\delta_{2i}\delta_{1j}$ we have
\begin{equation}
\dot{\sigma}_{12}=\dot{S}_{012}=[\alpha_1+\tau(\beta_1+\tau\gamma_1)+\tau]L_{21},
\end{equation}
and the associated shear moduli are $\alpha_1+\tau(\beta_1+\tau\gamma_1)$ and $\alpha_1+\tau(\beta_1+\tau\gamma_1)+\tau$,
respectively.

Thus, in total, there are six coefficients that are functions of $\tau$, but when linearized in $\tau$ there remain six constants,
namely $\alpha_1(0)$, $\alpha_2(0)$, $\beta_1(0)$, $\beta_2(0)$, $\alpha_1'(0)$ and $\alpha_2'(0)$, together with $\tau$.

%%%%%%%%%%%%%%%%%%%%%%%%%%%%%%%%%%%%%%%%%%

\subsection{Planar shear initial stress\label{sec5-3}}

%%%%%%%%%%%%%%%%%%%%%%%%%%%%%%%%%%%%%%%%%%

If the initial stress lies in the $(x_1,x_2)$ plane and is a pure shear stress of amount $\tau$ then
$\bm\tau=\tau(\mathbf{e}_1\otimes\mathbf{e}_2+\mathbf{e}_2\otimes\mathbf{e}_1)$ and the principal values of $\bm\tau$ are
$\pm\tau$. The principal axes of $\bm\tau$ bisect the background axes $\mathbf{e}_1$ and $\mathbf{e}_2$, along $\mathbf{\hat{e}}_{1,2}
= (\textbf{e}_1 \pm \textbf{e}_2)/\sqrt{2}$, say. Here, we therefore take as our axes of reference the principal axes
$\mathbf{\hat{e}}_1$, $\mathbf{\hat{e}}_2$, and $\mathbf{\hat{e}}_3=\mathbf{e}_3$; the associated components of $\bm{\mathcal{A}}_0$
are
\begin{eqnarray}
\hat{\mathcal{A}}_{01111}&=&2\alpha_1+\alpha_2+(1+4\beta_1+2\beta_2)\tau+(\beta_3+4\gamma_1+2\gamma_2)\tau^2+2\gamma_3\tau^3+\gamma_4\tau^4,\\[1ex]
\hat{\mathcal{A}}_{02222}&=&2\alpha_1+\alpha_2-(1+4\beta_1+2\beta_2)\tau+(\beta_3+4\gamma_1+2\gamma_2)\tau^2-2\gamma_3\tau^3+\gamma_4\tau^4,\\[1ex]
\hat{\mathcal{A}}_{03333}&=&2\alpha_1+\alpha_2,\quad \hat{\mathcal{A}}_{01122}=\alpha_2-\beta_3\tau^2+2\gamma_2\tau^2+\gamma_4\tau^4,\\[1ex]
\hat{\mathcal{A}}_{01133}&=&\alpha_2+\beta_2\tau+\gamma_2\tau^2,\quad \hat{\mathcal{A}}_{02233}=\alpha_2-\beta_2\tau+\gamma_2\tau^2,\\[1ex]
\hat{\mathcal{A}}_{01221}&=&\alpha_1+2\gamma_1\tau^2,\quad \hat{\mathcal{A}}_{01212}=\alpha_1+\tau +2\gamma_1\tau^2,\quad
\hat{\mathcal{A}}_{02121}=\alpha_1-\tau +2\gamma_1\tau^2,\\[1ex]
\hat{\mathcal{A}}_{03131}&=&\hat{\mathcal{A}}_{03113}=\alpha_1+\tau\beta_1+\tau^2\gamma_1,\quad
\hat{\mathcal{A}}_{01313}=\alpha_1+\tau+\tau\beta_1+\tau^2\gamma_1,\\[1ex]
\hat{\mathcal{A}}_{03232}&=&\hat{\mathcal{A}}_{03223}=\alpha_1-\tau\beta_1+\tau^2\gamma_1,\quad
\hat{\mathcal{A}}_{02323}=\alpha_1-\tau-\tau\beta_1+\tau^2\gamma_1.
\end{eqnarray}
Components referred to principal axes are indicated by a superposed hat.
We also have $\tr\bm\tau=\tr(\bm\tau^3)=0$, $\tr(\bm\tau^2)=2\tau^2$, and in general all the coefficients $\alpha_1,\dots,\gamma_4$ depend on $\tau$.

To illustrate the results in this case we consider the restriction to incremental plane strain with $\hat{L}_{3i}=
\hat{L}_{i3}=0$,\, $i=1,2,3$.  The components of the incremental nominal stress are then given by
\begin{equation}
\hat{\dot{S}}_{011}=\hat{\mathcal{A}}_{01111} \hat{L}_{11}+\hat{\mathcal{A}}_{01122} \hat{L}_{22},\quad
\hat{\dot{S}}_{022}=\hat{\mathcal{A}}_{01122} \hat{L}_{11}+\hat{\mathcal{A}}_{02222} \hat{L}_{22}
\end{equation}
for biaxial deformation parallel to the principal axes, and
\begin{equation}
\hat{\dot{S}}_{012}=(\alpha_1+2\gamma_1\tau^2)(\hat{L}_{12} + \hat{L}_{21}) +\tau \hat{L}_{21},\quad
\hat{\dot{S}}_{021}=(\alpha_1+2\gamma_1\tau^2)(\hat{L}_{12} + \hat{L}_{21}) - \tau \hat{L}_{12},
\end{equation}
for shearing deformations.

The plane strain Poisson's ratios and Young's modulus $E_1$ are then deduced as
\begin{equation}
\nu_{12}=\hat{\mathcal{A}}_{01122}/\hat{\mathcal{A}}_{02222},\quad \nu_{21}=\hat{\mathcal{A}}_{01122}/\hat{\mathcal{A}}_{01111},\quad
E_1=\hat{\mathcal{A}}_{01111}-\hat{\mathcal{A}}_{01122}^2/\hat{\mathcal{A}}_{02222}
\end{equation}
and, as in \eqref{nuEconnection}, $E_2=E_1\nu_{21}/\nu_{12}$. The shear moduli are $\alpha_1-\tau+2\gamma_1\tau^2$ and
$\alpha_1+\tau+2\gamma_1\tau^2$ for shear in the $\mathbf{\hat{e}}_1$ and $\mathbf{\hat{e}}_2$ directions, respectively.

When $\tau=0$ we recover the classical results for plane strain isotropy: Poisson's ratio is $\nu_{12}=\nu_{21}=\lambda_0/(\lambda_0+2\mu_0)$, and
Young's modulus is $E_1=E_2=4\mu_0(\lambda_0+\mu_0)/(\lambda_0+2\mu_0)$.

%%%%%%%%%%%%%%%%%%%%%%%%%%%%%%%%%%%%%%%%%%

\section{The effect of initial stress on infinitesimal wave propagation\label{sec6}}

%%%%%%%%%%%%%%%%%%%%%%%%%%%%%%%%%%%%%%%%%%

The theory of small amplitude (incremental) deformations or motions superimposed on a static finite deformation is well
established, but has received relatively little attention in the case of an initially stressed material with or without an
accompanying finite deformation except for works based on Biot's theory in the context of linear elasticity.  Here we consider
incremental motions in an infinite homogeneous medium subject to a homogeneous initial stress with $\mathcal{A}_{0piqj}$ having
the form given in \eqref{A0-comp-components-ref} and various specializations of that form.

From \eqref{motionhomogeneous} we recall that the equation of incremental motion is
\begin{equation}
\mathcal{A}_{0piqj}u_{j,pq}=\rho u_{i,tt}.
 \label{5motion-comp}
\end{equation}
Consider a homogeneous plane wave of the form
\begin{equation}
\mathbf{u}=\mathbf{m}f(\mathbf{n}\cdot\mathbf{x}- v t),
\end{equation}
where $\mathbf{m}$ is a fixed unit vector (the polarization vector), $f$ is a function of the argument
$\mathbf{n}\cdot\mathbf{x}- v t$ with appropriate regularity, $\mathbf{n}$ is a unit vector in the direction of propagation, and
$v$ is the wave speed. Substitution into the equation of motion \eqref{5motion-comp} (after dropping $f''$, which is assumed to
be non-zero) leads to
\begin{equation}
\mathcal{A}_{0piqj}n_pn_qm_j=\rho v^2 m_i.\label{prop-cond-compx}
\end{equation}
The associated \emph{acoustic tensor} $\mathbf{Q}(\mathbf{n})$ has components defined by
\begin{equation}
Q_{ij}(\mathbf{n})=\mathcal{A}_{0piqj}n_pn_q\label{acoustic-tensor}
\end{equation}
and enables the \emph{propagation condition} \eqref{prop-cond-compx} to be written compactly as
\begin{equation}
\mathbf{Q}(\mathbf{n})\mathbf{m}=\rho v^2\mathbf{m}.\label{prop-cond-comp}
\end{equation}

For any given direction of propagation $\mathbf{n}$ we have a three-dimensional symmetric algebraic eigenvalue problem for
determining $\rho v^2$ and $\mathbf{m}$. Because of the symmetry there are three mutually orthogonal eigenvectors $\mathbf{m}$
corresponding to the directions of displacement and the (three) values of $\rho v^2$ are obtained from the characteristic equation
\begin{equation}
\det[\mathbf{Q}(\mathbf{n})-\rho v^2 \mathbf{I}]=0.\label{secular}
\end{equation}
If $\mathbf{m}$ is known then $\rho v^2$ is given by
\begin{equation}
\rho v^2=[\mathbf{Q}(\mathbf{n})\mathbf{m}]\cdot\mathbf{m},
\end{equation}
and corresponds to a real wave speed provided $\rho v^2>0$, which is guaranteed if the \emph{strong ellipticity condition} holds, i.e. if
\begin{equation}
[\mathbf{Q}(\mathbf{n})\mathbf{m}]\cdot\mathbf{m}\equiv\mathcal{A}_{0piqj}n_pn_qm_im_j >0 \quad\mbox{for all non-zero vectors } \mathbf{m},\mathbf{n}.
\end{equation}
Then, a triad of waves with mutually orthogonal polarizations can propagate for any direction of propagation $\mathbf{n}$.
Henceforth we assume that the strong ellipticity condition holds. For detailed discussion of strong ellipticity we refer to \cite{True65} and \cite{Ogde84}, for example.

In the following we examine the effect of initial stress on the propagation of plane waves for the three examples of initial
stress considered in Sections \ref{sec4} and \ref{sec5}, and for this purpose we give explicit expressions for
$\mathbf{Q}(\mathbf{n})$ in each case.

%%%%%%%%%%%%%%%%%%%%%%%%%%%%%%%%%%%%%%%%%%

\subsection{Isotropic initial stress\label{sec6-1}}

%%%%%%%%%%%%%%%%%%%%%%%%%%%%%%%%%%%%%%%%%%

For the form of $\boldsymbol{\mathcal{A}}_0$ given by \eqref{isocalA}, with the connections \eqref{lambdaiso} and \eqref{muiso}
and \eqref{acoustic-tensor}, we obtain simply
\begin{eqnarray}
\mathbf{Q}(\mathbf{n}) =\mu(\tau)\mathbf{I}+[\lambda(\tau)+\mu(\tau)]\mathbf{n}\otimes\mathbf{n}.
 \label{Q-comp-ref}
\end{eqnarray}
As in the classical theory of isotropic elasticity with no initial stress there exists a longitudinal wave with speed $v_L$, say,
and two transverse waves with equal speeds $v_T$, say, for any direction of propagation.  With dependence on $\tau$ these are
given by
\begin{equation}
\rho v_L^2=\lambda(\tau)+2\mu(\tau),\quad \rho v_T^2=\mu(\tau).
\end{equation}
For sufficiently small initial stress we may linearize these expressions to give
\begin{align}
& \lambda(\tau)+2\mu(\tau) \simeq \lambda_0+2\mu_0+[2\alpha_1'(0)+\alpha_2'(0)+4\beta_1(0)+2\beta_2(0)+1]\tau,\label{lambda+2mu-iso}
\\[1ex]
& \mu(\tau) \simeq \mu_0+[\alpha_1'(0)+2\beta_1(0)+1]\tau,\label{mu-iso}
\end{align}
where $\alpha_1(0)=\mu_0,\alpha_2(0)=\lambda_0$.

Note that if, as in \cite{Tang67}, the isotropic moduli were defined based on the increment in the second Piola--Kirchhoff stress
according to \eqref{Tdotiso}  then $\mu(\tau)$ and $\lambda(\tau)+2\mu(\tau)$ would have to be replaced by $\mu(\tau)+\tau$ and
$\lambda(\tau)+2\mu(\tau)+\tau$, respectively, leading to the erroneous conclusion of \cite{Tang67} that when the elastic moduli
are independent of the initial stress the wave speeds are given by $\rho v_T^2=\mu_0-P$, $\rho v_L^2=\lambda_0+2\mu_0-P$ for the
case of a hydrostatic pressure ($\tau=-P$). Results such as these are not supported by the data shown in Tables \ref{table1} and \ref{table2}.

For a general isotropic elastic material under hydrostatic pressure \cite{True61} obtained expressions for the speeds of
longitudinal and transverse waves; see also \cite{True65}, section 75.

%%%%%%%%%%%%%%%%%%%%%%%%%%%%%%%%%%%%%%%%%%

\subsection{Uniaxial initial stress}
\label{uniaxial-propagation}
%%%%%%%%%%%%%%%%%%%%%%%%%%%%%%%%%%%%%%%%%%

For uniaxial initial stress the acoustic tensor is given by
\begin{equation}
\mathbf{Q}(\mathbf{n})=A\mathbf{I}+B\mathbf{a}\otimes\mathbf{a}+C\mathbf{n}\otimes\mathbf{n}+D(\mathbf{n}\otimes\mathbf{a}+\mathbf{a}\otimes\mathbf{n}),
\label{Quniaxial}
\end{equation}
where
\begin{align}
& A=\alpha_1+[1+\beta_1+\gamma_1\tau]\tau(\mathbf{n}\cdot\mathbf{a})^2,\notag\\[1ex]
& B=\beta_1\tau+\gamma_1\tau^2+[\beta_3+2\gamma_3\tau+\gamma_4\tau^2]\tau^2(\mathbf{n}\cdot\mathbf{a})^2,\notag\\[1ex]
& C = \alpha_1+\alpha_2,
\notag \\[1ex]
& D=[\beta_1+\beta_2+(\gamma_1+\gamma_2)\tau]\tau(\mathbf{n}\cdot\mathbf{a}).\label{ABCD}
\end{align}

If $\mathbf{n}=\mathbf{a}$ (propagation along the direction of unaxial stress), then there exists a longitudinal wave
with speed $v_{11}$ given by
\begin{equation}
\rho v_{11}^2=A+B+C+2D=2\alpha_1+\alpha_2+(1+4\beta_1+2\beta_2)\tau+(\beta_3+4\gamma_1+2\gamma_2)\tau^2
+2\gamma_3\tau^3+\gamma_4\tau^4,\label{uniaxlong}
\end{equation}
and two transverse waves with equal speeds $v_{12}$ given by
\begin{equation}
\rho v_{12}^2=A=\alpha_1+(1+\beta_1)\tau+\gamma_1\tau^2.\label{uniaxtrans}
\end{equation}
These formulas are consistent with the formulas (5.20) and (5.23) in \cite{Sham11} relating to propagation along a principal axis
of $\bm\tau$ except that in (5.20) there is a misprint (the coefficient of $\gamma_1$ should be 4 instead of 3 --- this arises
from the fact that in the expression (5.18) for $a$ in the latter paper $\gamma_1$ should be $2\gamma_1$). The linearized forms
of \eqref{uniaxlong} and \eqref{uniaxtrans} are
\begin{align}
& \rho v_{11}^2=\lambda_0+2\mu_0+[2\alpha_1'(0)+\alpha_2'(0)+4\beta_1(0)+2\beta_2(0)+1]\tau,\label{vLlinear}
\\[1ex]
& \rho v_{12}^2=\mu_0+[\alpha_1'(0)+\beta_1(0)+1]\tau,\label{vTlinear}
\end{align}
respectively.

On the other hand, if $\mathbf{n}\cdot\mathbf{a}=0$ (propagation transverse to the direction of uniaxial stress) then
\begin{equation}
A=\alpha_1,\quad B=\beta_1\tau +\gamma_1\tau^2,\quad C=\alpha_1+\alpha_2,\quad D=0,\label{orthogonal-case}
\end{equation}
and a longitudinal wave exists with speed $v_{22}$ given by $\rho v_{22}^2=A+C$.  There are also two transverse waves, with
polarizations along and perpendicular to $\mathbf{a}$ and speeds $v_{21}$ and $v_{23}$ given by
\begin{equation} \label{v21v23}
\rho v_{21}^2 = A+B = \alpha_1 + \beta_1\tau + \gamma_1\tau^2,
\quad
\rho v_{23}^2=A = \alpha_1,
\end{equation}
respectively.
We notice immediately that
\begin{equation}
\rho v_{12}^2 - \rho v_{21}^2 = \tau\label{v12v21}
\end{equation}
exactly, as expected from the property \eqref{property}. This well-known relationship (see, for example, \citealp{Biot65, Man87, Hoge93, Norr98})  forms the
basis of an experimental acoustic identification of a solid with anisotropy due to an initial stress, as opposed to a general
anisotropic linearly elastic solid without initial stress, for which $\rho v_{12}^2 = \rho v_{21}^2 = c_{66}$ (Voigt notation).  We note that, even earlier, \cite{Biot40}
obtained separate expressions for, in the present notation, $v_{12}$ and $v_{21}$ from which the above relationship may also be deduced.
Finally, we note that the linearized forms of $\rho v_{22}^2$ and \eqref{v21v23} are
\begin{equation}
\rho v_{22}^2=\lambda_0+2\mu_0+[2\alpha_1'(0)+\alpha_2'(0)]\tau,
\quad
\rho v_{21}^2=\mu_0+[\alpha_1'(0)+\beta_1(0)]\tau,
\quad
\rho v_{23}^2=\mu_0+\alpha_1'(0)\tau.\label{vTlinearortho}
\end{equation}

If the propagation takes place neither along the direction of uniaxial stress nor perpendicular to it, then several possibilities
arise, which are detailed in Appendix A.

%%%%%%%%%%%%%%%%%%%%%%%%%%%%%%%%%%%%%%%%%%

\subsection{Planar shear initial stress}

%%%%%%%%%%%%%%%%%%%%%%%%%%%%%%%%%%%%%%%%%%

Here we consider wave propagation in the plane of shear --- the $(x_1,x_2)$ plane.  The relevant components of the acoustic
tensor are then given by
\begin{eqnarray}
Q_{11}&=&\alpha_1+[\alpha_1+\alpha_2+2(2\gamma_1+\gamma_2)\tau^2+\gamma_4\tau^4]n_1^2+2(1+2\beta_1
+\beta_2+\gamma_3\tau^2)\tau n_1n_2 +(\beta_3+2\gamma_1)\tau^2n_2^2,\notag\\[1ex]
Q_{22}&=&\alpha_1+[\alpha_1+\alpha_2+2(2\gamma_1+\gamma_2)\tau^2+\gamma_4\tau^4]n_2^2+2(1+2\beta_1
+\beta_2+\gamma_3\tau^2)\tau n_1n_2 +(\beta_3+2\gamma_1)\tau^2n_1^2,\notag\\[1ex]
Q_{12}&=&(2\beta_1+\beta_2)\tau+\gamma_3\tau^3+[\alpha_1+\alpha_2+(2\gamma_1+2\gamma_2+\beta_3)\tau^2+\gamma_4\tau^4]n_1n_2,\notag
\end{eqnarray}
referred to background axes (not the principal axes considered in Section \ref{sec5-3}). As already noted, we assume that the
strong ellipticity condition holds so the wave speeds are real. As in the previous section we work in the $(x_1,x_2)$ plane with
$n_3=m_3=0$ and set $n_1=\cos\theta,n_2=\sin\theta$ and $m_1=\cos\phi,m_2=\sin\phi$. Then, by eliminating the wave speed from the
propagation condition, we may apply equation \eqref{propagationm1m2} from Appendix A, recast as
\begin{equation}
\tan2\phi=\frac{2(2\beta_1+\beta_2+\gamma_3\tau^2)\tau+[\alpha_1+\alpha_2+(2\gamma_1+2\gamma_2+\beta_3)\tau^2+\gamma_4\tau^4]\sin 2\theta}{
[\alpha_1+\alpha_2+(2\gamma_1+2\gamma_2-\beta_3+\gamma_4\tau^2)\tau^2]\cos 2\theta},\label{phithetashear}
\end{equation}
which gives $\phi$ for any given $\theta$.

The (in-plane) wave speeds are given by
\begin{equation}
\rho v^2=\frac{1}{2}\left[Q_{11}+Q_{22}\pm\sqrt{(Q_{11}-Q_{22})^2+4Q_{12}^2}\right].\label{speedsshear}
\end{equation}
For definiteness it is interesting to consider the situation in which $\tau$ is small and \eqref{phithetashear} is
linearized in $\tau$, which leads to
\begin{equation}
(\alpha_1+\alpha_2)\sin(2\phi-2\theta)=2(2\beta_1+\beta_2)\tau\cos 2\phi,
\end{equation}
from which we deduce that a longitudinal wave can propagate for $\tau\neq 0$ if either $\theta=\phi=\pi/4$ or
$2\beta_1+\beta_2=0$. The first of these possibilities corresponds to propagation along a principal axis and the second to a
special set of values of the material parameters that allows propagation of a longitudinal and transverse wave in any in-plane
direction.

In the linear specialization we obtain
\begin{align}
& Q_{11}+Q_{22} = 3\alpha_1+\alpha_2+2(1+2\beta_2+\beta_2)\tau \sin 2\theta,\\[1ex]
& (Q_{11}-Q_{22})^2+4Q_{12}^2 = (\alpha_1+\alpha_2)^2+4(\alpha_1+\alpha_2)(2\beta_1+\beta_2)\tau \sin 2\theta,
\end{align}
and the wave speeds \eqref{speedsshear} are then given by
\begin{eqnarray}
\rho v^2&=&\lambda_0+2\mu_0+\{2\alpha_1'(0)+\alpha_2'(0)+[1+4\beta_1(0)+2\beta_2(0)]\sin 2\theta\}\tau,\label{vLshearlinear}\\[1ex]
\rho v^2& =&\mu_0+[\alpha_1'(0)+\sin 2\theta]\tau,\label{vTshearlinear}
\end{eqnarray}
in which the coefficients have now be linearized in $\tau$.
These are respectively longitudinal and transverse when $\theta=\pi/4$, as indicated above, or in the special
case $2\beta_1+\beta_2=0$ the transverse wave speed stands but the longitudinal wave speed specializes accordingly.

\cite{Tang67} considered in-plane wave propagation for an initial shear stress in which the only non-zero components of the
second Piola--Kirchhoff stress were $T_{12}=T_{21}$.  As with the case of hydrostatic pressure discussed at the end of Section
\ref{sec6-1}, Tang used an incorrect incremental form of the constitutive law. When linearized in the initial stress the results from equation
(3.10) in his paper that parallel \eqref{vLshearlinear} and \eqref{vTshearlinear} can be shown to reduce, in the present notation, to
\begin{equation}
\rho v^2=\lambda_0+2\mu_0+(\lambda_0+3\mu_0)\bar{\tau}\sin 2\theta,\quad \rho v^2 =\mu_0,\label{shearlinearTang}
\end{equation}
where we have set $\bar{\tau}=\tau/\mu_0$ and $\tau=T_{12}$ since there is no distinction between stress measures themselves in
the reference configuration, which is quite different from the situation for their increments.  Note, in particular, that
\eqref{shearlinearTang}$_2$ depends on neither the initial shear stress nor the direction of propagation, which is quite
unrealistic and cannot be recovered from \eqref{vTshearlinear} for $\tau\neq 0$.  Fortuitously, \eqref{shearlinearTang}$_1$ can
be recovered from \eqref{vLshearlinear} by making the special choices $2\alpha_1'(0)+\alpha_2'(0)=0$ and
$2\beta_1(0)+\beta_2(0)=1+\lambda_0/2\mu_0$ of the coefficients.

%%%%%%%%%%%%%%%%%%%%%%%%%%%%%%%%%%%%%%%%%%

\subsection{Some connections with the classical theory of Biot}

%%%%%%%%%%%%%%%%%%%%%%%%%%%%%%%%%%%%%%%%%%

It is interesting now to consider how the classical theory of initial stress in the general linear theory of elasticity due to
Biot (see \citealp{Biot39,Biot40,Biot65}) relates to the present theory. As noted in Section \ref{sec1}, the elasticity tensor of
Biot, with components $\mathcal{B}_{piqj}$ depends in an unspecified way on the initially stressed configuration.  First, we
record that, as shown in \cite{Ogde11}, the general connection between $\mathcal{A}_{0piqj}$ and $\mathcal{B}_{piqj}$ may be
written in the form
\begin{equation}
\mathcal{A}_{0piqj} = \mathcal{B}_{piqj} - \tfrac{1}{2} \delta_{pj} \tau_{qi} - \tfrac{1}{2} \delta_{pq} \tau_{ij}
- \tfrac{1}{2} \delta_{qi} \tau_{pj} + \tfrac{1}{2}  \delta_{ij} \tau_{pq} + \delta_{qj} \tau_{pi},
\end{equation}
which can be shown to be equivalent to the expression (4.25) given in Chapter 2 of Biot's book (\citealp{Biot65}).
For the general expression \eqref{A0-comp-components-ref} to reduce to the Biot form for isotropic response, as quantified in
\eqref{biot-iso}, the material parameters in \eqref{A0-comp-components-ref} must be specialized to
\begin{equation}
\alpha_1=\mu_0, \quad\alpha_2=\lambda_0, \quad\beta_1=-1/2, \quad\beta_2=0,\label{biot1}
\end{equation}
where $\mu_0$ and $\lambda_0$ are the classical Lam\'e moduli and terms of order higher than 1 in $\bm\tau$ are neglected.

For the hydrostatic stress considered in Section 6.1 these specializations yield, from \eqref{mu-iso} and \eqref{lambda+2mu-iso},
$\rho v_T^2=\mu_0$ and $\rho v_L^2=\lambda_0+2\mu_0-\tau$.  The first of these agrees with the original result of Biot, who
mentioned that any dependence on the initial stress must be through the elastic constants themselves.  The results in the present
paper make the dependence explicit.  It does not appear that the result $\rho v_L^2=\lambda_0+2\mu_0-\tau$ was given by Biot.  

We remark that in \cite{Sham11} we adopted a slightly different form of the isotropic constitutive law from that given in \eqref{biot-iso}, namely
\begin{equation}
\mathcal{B}_{piqj}=\mu_0(\delta_{ij}\delta_{pq}+\delta_{qi}\delta_{pj})+\lambda_0\delta_{pi}\delta_{qj}
+\delta_{pi}\tau_{qj},\label{biot-iso2}
\end{equation}
for which, in the list \eqref{biot1}, $\beta_2=0$ is replaced by $\beta_2=1$. 

Turning next to the case of uniaxial stress we find from Section 6.2 first that for propagation in the direction of initial
stress the formulas \eqref{vLlinear} and \eqref{vTlinear} reduce to $\rho v_{11}^2=\lambda_0+2\mu_0-\tau$ and $\rho
v_{12}^2=\mu_0+\frac{1}{2}\tau$, respectively, while for propagation normal to the initial stress the formulas
\eqref{vTlinearortho} reduce to $\rho v_{22}^2=\lambda_0+2\mu_0$, $\rho v_{21}^2=\mu_0-\frac{1}{2}\tau$ and $\rho
v_{23}^2=\mu_0$.  For the case of compressive initial stress with $\tau=-P\,(P>0)$ the speeds of the relevant two transverse
waves agree with those obtained by Biot, specifically $\rho v_T^2=\mu_0\pm \frac{1}{2}P$.

It is also interesting to apply the Biot specialization to the shear stress example in Section 6.3.  Equations \eqref{vLshearlinear} and \eqref{vTshearlinear} yield
\begin{equation}
\rho v_L^2=\lambda_0+2\mu_0-\tau\sin2\theta,\quad \rho v_T^2=\mu_0+\tau\sin2\theta,
\end{equation}
and a longitudinal and transverse wave can propagate in any in-plane direction, where $\theta$ is the angle the propagation
direction makes with the principal direction of stress corresponding to principal stress $+\tau$.  Note that for propagation
along either principal direction there is no influence of $\tau$, which would seem to be unrealistic.

%%%%%%%%%%%%%%%%%%%%%%%%%%%%%%%%%%%%%%%%%%

\section{Deformed and pre-stressed isotropic elastic solid\label{sec7}}

%%%%%%%%%%%%%%%%%%%%%%%%%%%%%%%%%%%%%%%%%%

In this section we consider the initial stress to be associated with a finite deformation from an unstressed configuration (denoted $\mathcal{B}_0$) of an
isotropic elastic solid, and we shall consider two states of the accompanying stress analogous to those considered in the
foregoing sections, specifically pure dilatation, corresponding to hydrostatic stress, and an axially symmetric deformation
corresponding to uniaxial stress.  There is no direct analogue of the planar shear stress situation since when accompanied by
deformation such as simple shear there will in general also be normal stresses, which are not considered in Section \ref{sec4-3}.

For an isotropic elastic solid the Cauchy stress tensor $\bm\sigma$ is given by the appropriate specialization of \eqref{comp-sigma}, namely
\begin{equation}
J\bm\sigma =2 W_1 \mathbf{B} + 2 W_2 (I_1 \mathbf{B} - \mathbf{B}^2) + 2I_3W_3 \mathbf{I},\label{comp-sigma-iso}
\end{equation}
where we recall that $J=\det\mathbf{F}$ and $\mathbf{B}$ is the left Cauchy--Green tensor, where $\mathbf{F}$ and $J$ are
now measured relative to $\mathcal{B}_0$ instead of $\mathcal{B}_r$.  We may consider the strain energy to
depend on the three principal stretches $\lambda_1,\lambda_2,\lambda_3$ instead of the principal invariants $I_1,I_2,I_3$, and
for this purpose we write $W=\bar{W}(\lambda_1,\lambda_2,\lambda_3)$.  The principal Cauchy stresses $\sigma_1,\sigma_2,\sigma_3$
are then given simply by
\begin{equation}
J\sigma_1=\lambda_1\bar{W}_1,\quad J\sigma_2=\lambda_2\bar{W}_2,\quad J\sigma_3=\lambda_3\bar{W}_3,
\end{equation}
which are equivalent to the principal components of \eqref{comp-sigma-iso}, where $\bar{W}_i =\partial
\bar{W}/\partial\lambda_i,\, i=1,2,3$, and $J=\lambda_1\lambda_2\lambda_3$. This is easily seen by noting that in terms of the principal stretches
the invariants $I_1,I_2,I_3$ are given by
\begin{equation}
I_1=\lambda_1^2+\lambda_2^2+\lambda_3^2,\quad I_2=\lambda_2^2\lambda_3^2+\lambda_3^2\lambda_1^2+\lambda_1^2\lambda_2^2,
\quad I_3=J^2=\lambda_1^2\lambda_2^2\lambda_3^2.
\end{equation}

In terms of invariants the components of the elasticity tensor are given by
\begin{eqnarray}
J\mathcal{A}_{0piqj}&=&2(W_1+I_1W_2)B_{pq}\delta_{ij}+2W_2[2B_{pi}B_{qj}-B_{iq}B_{jp}-B_{pr}B_{rq}\delta_{ij}-B_{pq}B_{ij}]\notag\\[1ex]
&+&2I_3W_3(2\delta_{ip}\delta_{jq}-\delta_{iq}\delta_{jp})
+4W_{11}B_{ip}B_{jq}+4W_{22}(I_1 B_{ip}-B_{ir}B_{rp})(I_1 B_{jq}-B_{js}B_{sq})\notag\\[1ex]
&+&4W_{12}(2I_1B_{ip}B_{jq}-B_{ip}B_{jr}B_{rq}-B_{jq}B_{ir}B_{rp})+4I_3W_{13}(B_{ip}\delta_{jq}+B_{jq}\delta_{ip})
\notag\\[1ex]
&+&4I_3W_{23}[I_1(B_{ip}\delta_{jq}+B_{jq}\delta_{ip})-\delta_{ip}B_{jr}B_{rq}-\delta_{jq}B_{ir}B_{rp}]+4I_3^2W_{33}\delta_{ip}\delta_{jq},
\label{mathcalA0-components-comp}
\end{eqnarray}
where $B_{ij}$ are the components of $\mathbf{B}$.  This specializes the form of $J\mathcal{A}_{0piqj}$ for an initially-stressed
solid given in \cite{Sham11} to the present situation, but equivalent forms of \eqref{mathcalA0-components-comp} can be found in
the earlier literature on isotropic finite elasticity (see, for example, \citealp{Haye61}).

When referred to the principal axes of $\mathbf{B}$ the components \eqref{mathcalA0-components-comp} can be expressed more compactly as
\begin{equation}
J\mathcal{A}_{0iijj}=\lambda_i\lambda_j\bar{W}_{ij},\quad J\mathcal{A}_{0ijij}
=\frac{\lambda_i\bar{W}_i-\lambda_j\bar{W}_j}{\lambda_i^2-\lambda_j^2}\lambda_i^2,
\quad J\mathcal{A}_{0ijji}=J\mathcal{A}_{0ijij}-\lambda_i\bar{W}_i,\label{A-principal}
\end{equation}
where $\bar{W}_{ij}=\partial^2\bar{W}/\partial\lambda_i\partial\lambda_j,\, i,j\in\{1,2,3\}$. Note that the only non-zero
components of $\bm{\mathcal{A}}_0$ are $\mathcal{A}_{0iijj},\,i,j\in\{1,2,3\}$, together with, for $i\neq j$,
$\mathcal{A}_{0ijij}$ and $\mathcal{A}_{0ijji}$ (see, for example, \citealp{Ogde84}), as is the case for the components of the
elasticity tensor of an initially stressed but undeformed material when referred to the principal axes of the initial stress
$\bm\tau$, as can be seen by reference to \eqref{elasticity-tensor-principal-tau}.

If two of the principal stretches coincide, say $\lambda_j=\lambda_i$, then $\sigma_j=\sigma_i$ and a limiting process
can be applied to express the second and third elements in \eqref{A-principal} as
\begin{equation}
\mathcal{A}_{0ijij}\rightarrow\frac{1}{2}(\mathcal{A}_{0iiii}-\mathcal{A}_{0iijj}+\sigma_i),\quad
\mathcal{A}_{0ijji}\rightarrow\frac{1}{2}(\mathcal{A}_{0iiii}-\mathcal{A}_{0iijj}-\sigma_i),
\end{equation}
respectively.

%%%%%%%%%%%%%%%%%%%%%%%%%%%%%%%%%%%%%%%%%%

\subsection{Pure dilatation}

%%%%%%%%%%%%%%%%%%%%%%%%%%%%%%%%%%%%%%%%%%

Here we consider a pure dilatation with corresponding isotropic stress. Let $\lambda_i=J^{1/3},\, i=1,2,3,$ and define $\hat{W}(J)=\bar{W}(J^{1/3},J^{1/3},J^{1/3})$.  We denote the
corresponding (equal) principal stresses by $\sigma$ .  Then the simple
connection $\sigma=\hat{W}'(J)$ follows. The only independent components of $\bm{\mathcal{A}}_0$ are then
$\mathcal{A}_{0iijj}=J^{-1/3}\hat{W}_{ij}$ evaluated for the pure dilatation, with $\mathcal{A}_{0iiii}$ independent of $i$ and
$\mathcal{A}_{0iijj}$ independent of $i$ and $j\neq i$.  Now, from \eqref{Sdotsigmadotconnection}, we obtain
\begin{equation}
\mathbf{\dot{\bm\sigma}}=\bm{\mathcal{A}}_0\mathbf{L}+\sigma\mathbf{L}-\sigma(\tr\mathbf{L})\mathbf{I}.
\end{equation}
We consider an incremental simple shear deformation, and, without loss of generality, we may take this to be in the $(x_1,x_2)$ plane.  Then
\begin{equation}
\dot{\sigma}_{12}=\mathcal{A}_{01212}(L_{21}+L_{12})=\frac{1}{2}(\mathcal{A}_{0iiii}-\mathcal{A}_{0iijj}+\sigma)(L_{21}+L_{12}).
\end{equation}
This enables us to define the incremental shear modulus as a function of $J$, which we write as $\hat{\mu}(J)$.  It is given by
\begin{equation}
\hat{\mu}(J)=\frac{1}{2}(\mathcal{A}_{01111}-\mathcal{A}_{01122}+\sigma).\label{hatmu}
\end{equation}
Note that $\dot{S}_{021}=\mathcal{A}_{02121}L_{12}$ for simple shear in the $x_1$ direction,
$\dot{S}_{012}=\mathcal{A}_{01212}L_{21}$ for simple shear in the $x_2$ direction and $\mathcal{A}_{02121}=\mathcal{A}_{01212}$.
We note in passing that $\dot{T}_{012}=(\mathcal{A}_{01212}-\sigma)(L_{21}+L_{12})$ and it would be incorrect to define the shear
modulus as $\mathcal{A}_{01212}-\sigma$ based on use of $\mathbf{\dot{T}}_0$.

Next consider an incremental pure dilation $\varepsilon$, so that $L_{11}=L_{22}=L_{33}=\varepsilon/3$ and
\begin{equation}
\tr\mathbf{\dot{\bm\sigma}}=(\mathcal{A}_{01111}+2\mathcal{A}_{01122}-2\sigma)\varepsilon.
\end{equation}
The incremental bulk modulus, which we denote by $\hat{\kappa}(J)$, may be defined as a function of $J$ as
\begin{equation}
\hat{\kappa}(J)=\frac{1}{3}(\mathcal{A}_{01111}+2\mathcal{A}_{01122}-2\sigma),\label{hatkappa}
\end{equation}
where the components of $\bm{\mathcal{A}}_{0}$ are again evaluated for $\lambda_i=J^{1/3},\,i=1,2,3$. This may also be expressed
in the form $\hat{\kappa}(J)= J\hat{W}''(J)$, which agrees with a definition of the incremental bulk modulus adopted by
\cite{Scot07}.

If we set $\sigma=\tau$, where $\tau$ is the initial hydrostatic stress considered previously then, because of the connection
$\tau=\hat{W}'(J)$, we can in principle switch between the two different formulations, although, for a given $\tau$, this would
involve inversion of the relation $\tau=\hat{W}'(J)$ to obtain $J$.   When the switch is made we can identify $\hat{\mu}(J)$ and
$\hat{\kappa}(J)$ with $\mu(\tau)$ and $\kappa(\tau)$, respectively. We should note here that the densities in the stress-free
reference configuration $\mathcal{B}_0$, with density $\rho_0$, and the deformed (or initially-stressed) configuration $\mathcal{B}=\mathcal{B}_r$
are related by $\rho_0=\rho J$ and the factor $J$
needs to be used to switch between $\rho_0 v^2$ and $\rho v^2$ in considering formulas for various wave speeds, where $\rho=\rho_r$.

%%%%%%%%%%%%%%%%%%%%%%%%%%%%%%%%%%%%%%%%%%

\subsection{Uniaxial stretch with lateral contraction}

%%%%%%%%%%%%%%%%%%%%%%%%%%%%%%%%%%%%%%%%%%

Now consider a uniaxial stress $\sigma_1=\sigma$ with $\sigma_2=\sigma_3=0$ and stretches $\lambda_1$ and, by symmetry,
$\lambda_2=\lambda_3$. Then, $\bar{W}_2=\bar{W}_3=0$ and the components of $\bm{\mathcal{A}}_0$ are given by
\begin{align}
& J\mathcal{A}_{01111} = \lambda_1^2\bar{W}_{11},\quad J\mathcal{A}_{02222}=\lambda_2^2\bar{W}_{22}=J\mathcal{A}_{03333},
\quad J\mathcal{A}_{02233}=\lambda_2^2\bar{W}_{23},\\[1ex]
& J\mathcal{A}_{011ii} = \lambda_1\lambda_2\bar{W}_{12},\quad
J\mathcal{A}_{01i1i}=\frac{\lambda_1^3\bar{W}_1}{\lambda_1^2-\lambda_2^2},\quad
J\mathcal{A}_{0i1i1}=\frac{\lambda_1\lambda_2^2\bar{W}_1}{\lambda_1^2-\lambda_2^2},\quad i=2,3,\\[1ex]
& \mathcal{A}_{01ii1} = \mathcal{A}_{0i11i}=\mathcal{A}_{0i1i1}=\mathcal{A}_{01i1i}-\sigma,\quad i=2,3,\\[1ex]
& \mathcal{A}_{02323} = \mathcal{A}_{03232}=\mathcal{A}_{02332}=\mathcal{A}_{03223}=\frac{1}{2}(\mathcal{A}_{02222}-\mathcal{A}_{02233}),
\end{align}
all evaluated for $\lambda_3=\lambda_2$ and with $\sigma=\lambda_2^{-2}\bar{W}_1$. Poisson's ratios and Young's moduli can be
defined in exactly the same way as in Section 5.2, except that here the components of $\bm{\mathcal{A}}_0$ are different from
those in Section 5.2. There is no need to repeat them all here, but we note, for example, that
\begin{equation}
\nu_{12}=\mathcal{A}_{01122}/(\mathcal{A}_{02222}+\mathcal{A}_{02233})=\lambda_1\bar{W}_{12}/\lambda_2(\bar{W}_{22}+\bar{W}_{23})
\end{equation}
\begin{equation}
E_1=\mathcal{A}_{01111}-2\nu_{12}\mathcal{A}_{01122}=J^{-1}\lambda_1^2[\bar{W}_{11}-2\bar{W}_{12}^2/(\bar{W}_{22}+\bar{W}_{23})],
\end{equation}
which are both functions of $\lambda_1$ when $\lambda_2=\lambda_3$ is determined from $\bar{W}_2=0$ for a given form of
$\bar{W}$. The expression for $\nu_{12}$ is consistent with the definition of the incremental Poisson's ratio given by
\cite{Scot07}.  On the other hand, the definition of $E_1$ above differs from the corresponding definition in \cite{Scot07} since
the latter is defined, in the present notation, as $\lambda_1\mathrm{d}\sigma/\mathrm{d}\lambda_1$ with $\lambda_2=\lambda_3$
given by $\bar{W}_2=0$.  The definition of $E_1$ above corresponds to $\lambda_1\mathrm{d}\bar{W}_1/\mathrm{d}\lambda_1$ updated
to the deformed configuration by the push-forward factor $J^{-1}\lambda_1$, which is the appropriate specialization of the
general push forward operation $J^{-1}\mathbf{F}$ that takes the nominal stress $\mathbf{S}$ to the Cauchy stress
$\bm\sigma=J^{-1}\mathbf{F}\mathbf{S}$.  This definition of $E_1$ based on nominal stress is consistent with that used in Section 5.2.

%%%%%%%%%%%%%%%%%%%%%%%%%%%%%%%%%%%%%%%%%%

\subsection{Application to second-order elasticity}

%%%%%%%%%%%%%%%%%%%%%%%%%%%%%%%%%%%%%%%%%%

For definiteness we now specialize the form of strain-energy function $W$ to the third order in the strain.  The precise form of
this approximation depends on the choice of strain measure, but here, in order to make contact with several contributions to the
literature, we shall use the Green strain tensor $\mathbf{E}=\frac{1}{2}(\mathbf{C}-\mathbf{I})$, where $\mathbf{C}$ is again the
right Cauchy--Green deformation tensor.  The aim here is to obtain the first order
correction to the classical longitudinal and transverse wave speeds. For a discussion of the advantages of using logarithmic strain
instead of Green strain, particularly when approaching the incompressible limit, we refer to the recent paper by \cite{Dest10}.

%%%%%%%%%%%%%%%%%%%%%%%%%%%%%%%%%%%%%%%%%%

\subsubsection{Historical expansions of the strain energy}

%%%%%%%%%%%%%%%%%%%%%%%%%%%%%%%%%%%%%%%%%%

The use of invariants for expressing the third-order expansion of $W$ appears to have been introduced by \cite{Bril25}; see also
his monograph \citep{Bril46}.  The Brillouin expansion may be written in the form
\begin{equation}
W=W_0-\frac{1}{2}p_0\tilde{I}_1+\frac{1}{8}\lambda_0\tilde{I}_1^2+\frac{1}{4}\mu_0\tilde{I}_2+A\tilde{I}_1\tilde{I}_2
+B\tilde{I}_1^3+C\tilde{I}_3,
\end{equation}
where, in different notation from that used by Brillouin, $\tilde{I}_1=2\tr\mathbf{E}$, $\tilde{I}_2=4\tr(\mathbf{E}^2)$,
$\tilde{I}_3=8\tr(\mathbf{E}^3)$, respectively of orders 1, 2, 3 in $\mathbf{E}$, the constant $W_0$ is the energy in the
reference configuration and $p_0$ corresponds to an initial pressure in the reference configuration.  Note, in particular, that
Brillouin used $2\mathbf{E}$ rather than $\mathbf{E}$ itself as the strain measure and that the notations $A,B,C$ are different from
those defined in \eqref{ABCD}.

Let us drop the terms $W_0$ and $p_0$, which are redundant for our purposes, and recast the remaining terms using the principal
invariants of Green strain, which, for consistency with the notation used in \cite{Dest10}, we denote by $i_1,i_2,i_3$. Thus,
\begin{equation}
i_1=\tr\mathbf{E},\quad i_2=\frac{1}{2}[i_1^2-\tr(\mathbf{E}^2)],\quad i_3=\det\mathbf{E}.
\end{equation}
Then, we have
\begin{equation}
W=\frac{1}{2}(\lambda_0+2\mu_0)i_1^2-2\mu_0 i_2+8(A+B+C)i_1^3-8(2A+3C)i_1i_2+24C i_3.
\end{equation}
This is entirely equivalent to the strain-energy function generally referred to as the Murnaghan form of strain energy that
appears in \cite{Murn51,Murn67} and is based on the use of Green strain. It is written
\begin{equation}
W=\frac{1}{2}(\lambda_0+2\mu_0)i_1^2-2\mu_0 i_2+\frac{1}{3}(l+2m)i_1^3-2m i_1i_2+ni_3,\label{Murnaghan-Green}
\end{equation}
where $l,m,n$ are the Murnaghan constants.  The Brillouin constants $A,B,C$ are given in terms of $l,m,n$ by
\begin{equation}
A=\frac{1}{16}(2m-n),\quad B=\frac{1}{48}(2l-2m+n),\quad C=\frac{1}{24}n.
\end{equation}

We remark that in his original paper \cite{Murn37} worked in terms of the Almansi strain tensor $(\mathbf{I}-\mathbf{B}^{-1})/2$
and its principal invariants, which we denote here by $\bar{I}_1,\bar{I}_2,\bar{I}_3$.  The original energy function of Murnaghan
has the form
\begin{equation}
W=\frac{1}{2}(\lambda_0+2\mu_0)\bar{I}_1^2-2\mu_0 \bar{I}_2+\bar{l}\bar{I}_1^3+\bar{m} \bar{I}_1\bar{I}_2
+\bar{n}\bar{I}_3.\label{Murnaghan-Almansi}
\end{equation}
In general \eqref{Murnaghan-Almansi} is different from \eqref{Murnaghan-Green}, but the two are equivalent to the third order in
the strains.  It is then easy to show that the constants $l,m,n$ and $\bar{l},\bar{m},\bar{n}$ are related by
\begin{equation}
\bar{l}=\frac{1}{3}(l+2\lambda_0),\quad \bar{m}=-2m-4\lambda_0-12\mu_0,\quad \bar{n}=n+12\mu_0.
\end{equation}
This difference has significance when considering approximations to wave speeds at this order, as we shall see shortly. We note
in passing that because of the requirement of objectivity the Almansi strain tensor (or any other Eulerian strain tensor) can be
used as the argument of the strain energy function if and only if the material is isotropic.  Clearly, the work of Brillouin on
this topic has been to some extent overlooked, although \cite{Murn37}, \cite{HuKe53}, and \cite{True61} did refer to
\cite{Bril25}.  For further historical discussion of second-order elasticity, including the contribution of \cite{Rivl53}, we
refer to section 66 of \cite{True65}.

An equivalent form of the third-order expanded energy function was also introduced by \cite{Land37}, who were apparently unaware
of the work of Brillouin.  This may be written
\begin{equation}
W=\frac{1}{2}\lambda_0 (\tr\mathbf{E})^2+\mu_0 \tr(\mathbf{E}^2)+\frac{1}{3}\bar{A}\,\tr(\mathbf{E}^3)
+\bar{B}(\tr\mathbf{E})\tr(\mathbf{E}^2)+\frac{1}{3}\bar{C}(\tr\mathbf{E})^3,
\end{equation}
where overbars have been used to distinguish the third-order constants from those of Brillouin; in \cite{Land86} the notations
$A,B,C$ were used, differing from those in \cite{Land37}.  The connections between the Murnaghan constants $l,m,n$ and the Landau
constants $\bar{A},\bar{B},\bar{C}$ were noted in \cite{Dest10} as
\begin{equation}
\bar{A}=n,\quad \bar{B}=m-\frac{1}{2}n,\quad \bar{C}=l-m+\frac{1}{2}n.
\end{equation}

\cite{Biot40x} also developed a third-order expansion, which is equivalent to the above and details may also be found in his book
\citep{Biot65}.  \cite{Biot65} worked in terms of the principal strain components $\lambda_i-1$ and the corresponding principal
Biot stresses.  His third-order constants, denoted $D,F,G$, can be shown to be related to $l,m,n$ via
\begin{equation}
D=l+2m +\frac{3}{2}(\lambda_0+2\mu_0), \quad F=l+\frac{1}{2}\lambda_0,\quad G=2l-2m+n.
\end{equation}
He did not give the form of strain energy explicitly.

Finally, we mention the third-order expansion adopted by \cite{Toup61}, who used the invariants $\tr\mathbf{E}$,
$\tr(\mathbf{E}^2)$, $\tr(\mathbf{E}^3)$ and third-order constants $\nu_1,\nu_2,\nu_3$.  In terms of the principal invariants of
Green strain their energy function has the form
\begin{equation}
W=\frac{1}{2}(\lambda_0+2\mu_0)i_1^2-2\mu_0 i_2+\frac{1}{6}(\nu_1+6\nu_2+8\nu_3)i_1^3-2(\nu_2+2\nu_3)i_1i_2+4\nu_3 i_3.
\end{equation}
It is straightforward to show that the constants $\nu_1,\nu_2,\nu_3$ are related to $l,m,n$ and $\bar{A},\bar{B},\bar{C}$ by \citep{Norr98}
\begin{equation}
\nu_1=2l-2m+n=2\bar{C},\quad \nu_2=m-\frac{1}{2}n=\bar{B},\quad \nu_3=\frac{1}{4}n=\frac{1}{4}\bar{A}.
\end{equation}
Connections between some of the above sets of constants and others used by \cite{Rivl53} were also noted by \cite{True65}.

Here we shall work in terms of the Murnaghan constants $l,m,n$ but cast \eqref{Murnaghan-Green} in terms of the principal
invariants \eqref{invs1} of $\mathbf{C}$ as
\begin{multline}
W =
\dfrac{\lambda_0}{8} (I_1 - 3)^2 + \dfrac{\mu_0}{4}(I_1^2 - 2I_1 - 2I_2 + 3)
\\[1ex]
+ \dfrac{l}{24}(I_1 - 3)^3
+ \dfrac{m}{12}(I_1 - 3)(I_1^2 - 3I_2) + \dfrac{n}{8}(I_1 - I_2 + I_3 - 1).
\label{Murnaghan}
\end{multline}

For this energy function we have $W_{22}=W_{13}=W_{23}=W_{33}=0$ and the expression \eqref{mathcalA0-components-comp} reduces to
\begin{eqnarray}
J\mathcal{A}_{0piqj}
&=&2(W_1+I_1W_2)B_{pq}\delta_{ij}+2W_2[2B_{pi}B_{qj}-B_{iq}B_{jp}-B_{pr}B_{rq}\delta_{ij}-B_{pq}B_{ij}]
\notag\\[1ex]
 &+& 2I_3W_3(2\delta_{ip}\delta_{jq}-\delta_{iq}\delta_{jp})+4W_{11}B_{ip}B_{jq}\notag\\[1ex]
&+& 4W_{12}(2I_1B_{ip}B_{jq}-B_{ip}B_{jr}B_{rq}-B_{jq}B_{ir}B_{rp}),
\end{eqnarray}
and the remaining coefficients $W_1$, $W_2$, $W_3$, $W_{11}$ and $W_{12}$ are simply obtained from \eqref{Murnaghan}.

In working with second-order elasticity the corrections to the classical moduli are obtained by expanding the coefficients in the
above to the first order in $\mathbf{E}$.  We have $\mathbf{C}=\mathbf{I}+2\mathbf{E}$ and $I_1=3+2E$, exactly, which we use
together with the linear approximations $I_2 \simeq 3+4E$, $I_3 \simeq 1 +2E$, $\mathbf{B} \simeq
\mathbf{C}=\mathbf{I}+2\mathbf{E}$, where $E=\tr\mathbf{E}$.  We also note that $\rho \simeq \rho_r(1-E)$. To the first order in
$\mathbf{E}$ we obtain
\begin{eqnarray}
W_1&=&
\mu_0+\tfrac{1}{8}n+\tfrac{1}{2}(\lambda_0+2\mu_0+2m)E,\quad W_2=-\tfrac{1}{2}\mu_0-\tfrac{1}{8}n-\tfrac{1}{2}mE,\quad W_3=\tfrac{1}{8}n,
\notag \\[1ex]
W_{11}
&=& \tfrac{1}{4}(\lambda_0+2\mu_0+4m)+\tfrac{1}{2}(l+2m)E,\quad W_{12}=-\tfrac{1}{4}m,
\end{eqnarray}
and hence
\begin{eqnarray}
J\mathcal{A}_{0piqj}&\simeq&\mu_0(\delta_{ij}\delta_{pq}+\delta_{iq}\delta_{jp})+\lambda_0\delta_{ip}\delta_{jq}
+\frac{1}{2}[(2\lambda_0+2m-n)\delta_{ij}\delta_{pq}+2(2l-2m+n)\delta_{ip}\delta_{jq}\notag\\[1ex]
&+&(2m-n)\delta_{iq}\delta_{jp}]E+\frac{1}{2}(4\mu_0+n)(\delta_{pq}E_{ij}+\delta_{iq}E_{jp}+\delta_{jp}E_{iq})\notag\\[1ex]
&+&\frac{1}{2}(8\mu_0+n)\delta_{ij}E_{pq} +(2\lambda_0+2m-n)(\delta_{ip}E_{jq}+\delta_{jq}E_{ip}).\label{secondorderA}
\end{eqnarray}

We are particularly interested in the case of a pure dilatation, for which, with $E_{ij}=E\delta_{ij}/3$,
\eqref{secondorderA} reduces to
\begin{eqnarray}
J\mathcal{A}_{0piqj}
&=&
\mu_0(\delta_{ij}\delta_{pq}+\delta_{iq}\delta_{jp})+\lambda_0\delta_{ip}\delta_{jq}+\frac{1}{6}(12\mu_0 +6\lambda_0 +6m-n)E\delta_{ij}\delta_{pq}
\notag\\[1ex]
&+&\frac{1}{6}(8\mu_0+6m-n)E\delta_{iq}\delta_{pj}+
\frac{1}{3}(4\lambda_0+6l-2m+n)E\delta_{pi}\delta_{qj}.\label{dilatationAsecondorder}
\end{eqnarray}

From the definitions \eqref{hatmu} and \eqref{hatkappa} we now obtain the approximations
\begin{equation}
\hat{\mu}(J)\simeq \mu_0+\frac{1}{6}(6\lambda_0+6\mu_0+6m-n)E
\end{equation}
and
\begin{equation}
\hat{\kappa}(J)\simeq \kappa_0+\frac{2}{9}(9l+n)E,
\end{equation}
for small dilatation, where to the first order $J=1+E$, and we note that $\hat{\mu}(1)=\mu_0$, $\hat{\kappa}(1)=\kappa_0$.

To the second order we may expand the isotropic stress as
\begin{equation}
\sigma=\hat{W}'(J)\simeq \varepsilon \hat{W}''(1)+\frac{1}{2}\varepsilon^2\hat{W}'''(1),
\end{equation}
where $\hat{W}''(1)=\kappa_0$, $\varepsilon\equiv J-1\simeq E+E^2/6$ and, for the Murnaghan strain energy, $\hat{W}'''(1)=-\kappa_0+2l+2n/9$.

As indicated earlier, if the initial stress $\bm\tau$ discussed in Sections \ref{sec4} and \ref{sec5} is associated with a pure
dilatation, so that $\bm\tau=\tau\mathbf{I}$, then in principle the results here can be converted to those based on $\tau$.  In particular, if we set
$\sigma=\tau$ in the above and work to second order then we may invert the $\tau\longleftrightarrow\varepsilon$ relation in the
form
\begin{equation}
\varepsilon\simeq\tau/\kappa_0-\frac{1}{2}\hat{W}'''(1)\tau^2/\kappa_0^3,
\end{equation}
but in considering the linear approximations of the shear and bulk moduli only the first order term need be retained. Then, with
$\tau=\kappa_0\varepsilon$, we can identify $\mu(\tau)$ and $\kappa(\tau)$ with $\hat{\mu}(J)$ and $\hat{\kappa}(J)$,
respectively.  Thus
\begin{align}
& \mu(\tau)\simeq \mu_0 +[\alpha_1'(0)+2\beta_1(0)+1]\kappa_0\varepsilon\simeq\hat{\mu}(J)\simeq \mu_0
+\frac{1}{6}(6\lambda_0+6\mu_0+6m -n)\varepsilon,
\notag \\[1ex]
& \kappa(\tau)\simeq\kappa_0+\frac{1}{3}[2\alpha_1'(0)+4\beta_1(0)+3\alpha_2'(0)+6\beta_2(0)-1]\kappa_0\varepsilon\simeq\hat{\kappa}(J)
\simeq\kappa_0+\frac{2}{9}(9l+n)\varepsilon.
\end{align}
Hence we can relate the constants $\alpha_1'(0)$, $\alpha_2'(0)$, $\beta_1(0)$, $\beta_2(0)$ to the Murnaghan constants. After a
little rearrangement, this yields
\begin{eqnarray}
\alpha_1'(0)+2\beta_1(0)&=&(2\mu_0+6m-n)/6\kappa_0,\label{a1primeb1}\\[1ex]
\alpha_2'(0)+2\beta_2(0)&=&(\lambda_0+6l-2m+n)/3\kappa_0.\label{a2primeb2}
\end{eqnarray}

In fact, the separate values of $\alpha_1'(0)$, $\alpha_2'(0)$, $\beta_1(0)$ and $\beta_2(0)$ can be obtained by considering other initial deformations than pure dilatation, and we show this
in Section 7.3.3 by comparing wave speeds based on the theory of uniaxial initial stress from Section 6.2 with those based on a finite deformation from an isotropic reference configuration under uniaxial stress.

%%%%%%%%%%%%%%%%%%%%%%%%%%%%%%%%%%%%%%%%%%

\subsubsection{Implications for the wave speeds}

%%%%%%%%%%%%%%%%%%%%%%%%%%%%%%%%%%%%%%%%%%

Turning now to the propagation of plane waves, we note that for any ($\mathbf{m}$, $\mathbf{n}$) pair satisfying \eqref{prop-cond-comp} the wave speed $v$ is given by
\begin{equation}
\rho_0 v^2= J[\mathbf{Q}(\mathbf{n})\mathbf{m}]\cdot\mathbf{m}=J\mathcal{A}_{0piqj}n_pn_qm_im_i,\label{vQAconnections}
\end{equation}
where the factor $J$ is now included to reflect the change in reference configuration from $\mathcal{B}_0$ to $\mathcal{B}_r=\mathcal{B}$.
For the case of pure dilatation we then obtain, on use of \eqref{dilatationAsecondorder},
\begin{multline}
\rho_0 v^2
=
\mu_0 +(\mu_0+\lambda_0)(\mathbf{m}\cdot\mathbf{n})^2
+ \frac{1}{6}(12\mu_0 +6\lambda_0+6m-n)E\\[1ex]
+\frac{1}{6}[8\lambda_0+8\mu_0+12l+2m+n]E(\mathbf{m}\cdot\mathbf{n})^2.
\end{multline}
For a longitudinal wave with $\mathbf{m}=\mathbf{n}$ this reduces to
\begin{equation}
\rho_0 v_L^2=\lambda_0+2\mu_0+\frac{1}{3}(7\lambda_0+10\mu_0+6l+4m)E,
\end{equation}
and for a transverse wave with $\mathbf{m}\cdot\mathbf{n}=0$
\begin{equation}
\rho_0 v_T^2=\mu_0+(\lambda_0 + 2\mu_0 + m - \frac{1}{6}n)E.
\end{equation}
These results agree with those obtained by \cite{HuKe53} for the case of hydrostatic pressure ($\tau=-P$). They used the
Murnaghan energy function based on Green strain. As shown by \cite{Sham11} there is similar agreement in the case of the uniaxial
compression considered by \cite{HuKe53}.  \cite{Toup61} obtained equivalent results based on their third-order expansion, but
expressed in terms of the acoustoelastic coefficients, which we write as
\begin{align}
&\frac{\text{d}}{\text{d}E}(\rho_0 v_L^2)\big\vert_{E=0}=\frac{1}{3}(7\lambda_0+10\mu_0+3\nu_1+10\nu_2+8\nu_3),
\\[1ex]
&\frac{\text{d}}{\text{d}E}(\rho_0 v_T^2)\big\vert_{E=0}=\frac{1}{3}(3\lambda_0+6\mu_0+3\nu_2+4\nu_3).
\end{align}
Note that the $E$ in \cite{Toup61} is $3\times$ that used here. \cite{Toup61} mentioned that their results were equivalent to
those obtained by \cite{Bril25}.  The Brillouin results therefore pre-date much of the subsequent work.

In calculating the second order correction to longitudinal and transverse wave speeds, \cite{Birc38} used the original (Almansi
strain based) form of the Murnaghan strain energy but effectively set $\bar{l}=\bar{m}=\bar{n}=0$, obtaining the results
\begin{align}
&\rho_0 v_L^2=\lambda_0+2\mu_0+P(13\lambda_0+14\mu_0)/(3\kappa_0)\equiv \lambda_0+2\mu_0-(13\lambda_0+14\mu_0)E/3, \\[1ex]
& \rho_0 v_T^2=\mu_0+P(\lambda_0+2\mu_0)/\kappa_0\equiv \mu_0-(\lambda_0+2\mu_0)E\label{Birchtrans}
\end{align}
for the wave speeds, where $P=-\kappa_0E$ is the pressure.  The omission of the third-order constants can be significant since
typically they are of the same order of magnitude as the Lam\'e moduli $\lambda_0$ and $\mu_0$, as illustrated by the data shown
in Table \ref{table3}.

\begin{table}[!h]
\caption{\textit{Lam\'e constants and Landau third-order elastic moduli for five solids (expressed in units of }$10^{9}$\,N\,m$^{-2}$\textit{), as collected by
\cite{Poru03}: his Murnaghan constants $m$, $n$, $l$ have been converted here to the Toupin--Bernstein constants $\nu_1$,
$\nu_2$, $\nu_3$}}
\begin{center}
\begin{tabular}{lrrrrr}
\hline\noalign{\smallskip}
Material & $\lambda_0$ & $\mu_0$ & $\nu_1$ & $\nu_2$ & $\nu_3$ \\
\noalign{\smallskip}\hline\noalign{\smallskip}
      Polystyrene & $ 1.71$ & $-21.2$ & $-10$ & $-8.3$ & $-2.5$ \\
            Steel Hecla 37 & $-354$ & $82.1$ & $-358$ & $-282$ & $-89.5$ \\
            Aluminium 2S  & $-204$ & $27.6$ & $-228$ & $-197$ & $-57$ \\
            Pyrex glass & $264$ & $27.5$ & $420$ & $-118$ & $105$ \\
            SiO$_2$ melted & $72$ & $31.3$ & $-44$ & $93$ & $-11$ \\
            \noalign{\smallskip}\hline
\end{tabular}
\end{center}
\label{table3}
\end{table}

If the third-order constants are retained then we have, instead,
\begin{align}
& \rho_0 v_L^2=\lambda_0+2\mu_0-(13\lambda_0+14\mu_0-18\bar{l}+2\bar{m})E/3,
\notag \\[1ex]
& \rho_0 v_T^2=\mu_0-(\lambda_0+2\mu_0+\frac{1}{2}\bar{m}+\frac{1}{6}\bar{n})E.
\end{align}

If, by contrast, we set the third-order constants $l,m,n$ to zero then the \cite{HuKe53} results reduce to
\begin{align}
& \rho_0 v_L^2=\lambda_0+2\mu_0+(7\lambda_0+10\mu_0)E/3,
\notag \\[1ex]
&
\rho_0 v_T^2=\mu_0+(\lambda_0+2\mu_0)E.
\end{align}
Note, in particular, the opposite sign but equal magnitude of the second term in the shear wave expression compared with
\eqref{Birchtrans}.  Thus, interpretation of the results requires caution.  In particular, care must be taken that the results
allow for an increase as well as for a decrease of the wave speeds with pressure and uniaxial stress, depending on which solid is
considered (see Tables \ref{table1} and \ref{table2}).

%%%%%%%%%%%%%%%%%%%%%%%%%%%%%%%%%%%%%%%%%%

\subsubsection{The case of uniaxial stress}

%%%%%%%%%%%%%%%%%%%%%%%%%%%%%%%%%%%%%%%%%%

We now consider a deformation from a stress-free configuration $\mathcal{B}_0$ of an isotropic material associated with a uniaxial stress $\sigma$ in the $\mathbf{e}_1$ direction.
We denote the corresponding component $E_{11}$ of the Green strain tensor by $E$.  Then, by setting the lateral stress to zero and by symmetry, $E_{22}=E_{33}=-\lambda_0E/2(\lambda_0+\mu_0)$, to the first order in $E$, and $\sigma=3\kappa_0\mu_0E/(\lambda_0+\mu_0)$, and hence
\begin{equation}
J=1+\mu_0E/(\lambda_0+\mu_0)=1+\sigma/3\kappa_0,
\end{equation}
also to first order.

The wave speeds $v_{11},v_{12},v_{22},v_{23}$ are then calculated by using \eqref{secondorderA} and appropriate specializations of \eqref{vQAconnections} and the connection
$\rho_0=\rho J$.  After some manipulations, which are omitted, this yields the formulas
\begin{eqnarray}
\rho v_{11}^2&=&\lambda_0+2\mu_0+2[2\lambda_0^2+7\lambda_0\mu_0+4\mu_0^2+\mu_0 l+2(\lambda_0+\mu_0)m]\sigma/3\kappa_0\mu_0,\\[1ex]
\rho v_{12}^2&=&\mu_0+[4\mu_0(4\lambda_0+3\mu_0)+4\mu_0 m+\lambda_0 n]\sigma/12\kappa_0\mu_0,\\[1ex]
\rho v_{22}^2&=&\lambda_0+2\mu_0-(2\lambda_0^2+5\lambda_0\mu_0+2\mu_0^2-2\mu_0 l+2\lambda_0 m)\sigma/3\kappa_0\mu_0,\\[1ex]
\rho v_{23}^2&=&\mu_0-[2\mu_0(2\lambda_0+\mu_0)-2\mu_0 m+(\lambda_0+\mu_0)n]\sigma/6\kappa_0\mu_0,
\end{eqnarray}
and we also note the connection $\rho v_{12}^2-\rho v_{21}^2=\sigma$.
By setting $\sigma=\tau$ we then compare these results with the formulas in \eqref{vLlinear}, \eqref{vTlinear} and \eqref{vTlinearortho}, which we now collect together as
\begin{eqnarray}
\rho v_{11}^2&=&\lambda_0+2\mu_0+[2\bar{\alpha}_1'(0)+\bar{\alpha}_2'(0)+4\beta_1(0)+2\beta_2(0)+1]\tau,\\[1ex]
\rho v_{12}^2&=&\mu_0+[\bar{\alpha}_1'(0)+\beta_1(0)+1]\tau,\\[1ex]
\rho v_{22}^2&=&\lambda_0+2\mu_0+[2\bar{\alpha}_1'(0)+\bar{\alpha}_2'(0)]\tau,\\[1ex]
\rho v_{23}^2&=&\mu_0+\bar{\alpha}_1'(0)\tau,
\end{eqnarray}
with $\rho v_{21}^2$ given by \eqref{v12v21}.  Note, in particular, that bars have now been placed over $\alpha_1'(0)$ and $\alpha_2'(0)$.
This is because the arguments of $\alpha_1$ and $\alpha_2$ are different for hydrostatic stress and uniaxial stress, respectively the relevant
invariants of $\bm\tau$ are $(3\tau,3\tau^2,3\tau^3)$ and
$(\tau,\tau^2,\tau^3)$, so that $\alpha_1'(0)=3\bar{\alpha}_1'(0)$ and $\alpha_2'(0)=3\bar{\alpha}_2'(0)$, while $\beta_1(0)$ and $\beta_2(0)$ are the same in each case.

Comparison the two sets of formulas yields the results
\begin{eqnarray}
\beta_1(0)&=&1+n/4\mu_0,\quad \beta_2(0)=(2\lambda_0+2m-n)/2\mu_0,\\[1ex]
\alpha_1'(0)&=&-[2(2\lambda_0+\mu_0)\mu_0-2\mu_0 m+(\lambda_0+\mu_0)n]/2\kappa_0\mu_0,\\[1ex]
\alpha_2'(0)&=&-[(2\lambda_0+\mu_0)\lambda_0-2\mu_0l+(2m-n)(\lambda_0+\mu_0)]/\kappa_0\mu_0,
\end{eqnarray}
from which it is easy to check that the results \eqref{a1primeb1} and \eqref{a2primeb2} are recovered.

It is interesting that the \emph{four} constants $\alpha_1'(0)$, $\alpha_2'(0)$, $\beta_1(0)$ and $\beta_2(0)$ are expressed in terms of the \emph{three} Murnaghan constants.
This is explained by noting that the anisotropic constitutive law for an initially stressed material with no accompanying deformation is specialized to isotropy by introducing a stress-free reference configuration and an associated deformation.

Finally, we note that the Biot values \eqref{biot1} are obtained from the latter formulas by specializing the Murnaghan constants to $m=-2l=-(\lambda_0+2\mu_0),\, n=-6\mu_0$.

%%%%%%%%%%%%%%%%%%%%%%%%%%%%%%%%%%%%%%%%%%

\section{Concluding remarks}

%%%%%%%%%%%%%%%%%%%%%%%%%%%%%%%%%%%%%%%%%%

In this paper we have examined in detail the effect of initial stress on the propagation of small amplitude homogeneous plane
waves in an undeformed elastic material on the basis of the general theory of a hyperelastic material with initial stress developed by \cite{Sham11},
which had its genesis in the work of \cite{Hoge85,Hoge86,Hoge93,Hoge93b} in particular. A key feature of the constitutive law,
formulated in terms of invariants of the deformation and initial stress, is that the elasticity tensor depends in general in a
highly nonlinear way on initial stress.  Important special cases considered within the general framework included initial
stresses corresponding to hydrostatic stress, uniaxial stress, and shear stress, for which explicit and relatively simple forms
of the elasticity tensor were given.  For each of these states of stress the dependence of various elastic moduli on the initial
stress was made explicit.  For example, simple formulas were obtained for the stress-dependence of the Lam\'e moduli in the case of isotropic
initial stress and Poisson's ratios and Young's moduli for the cases of uniaxial initial stress and planar initial shear stress.

The results were applied to infinitesimal wave propagation and it was shown how some known results fit within the general
framework, and some discrepancies in some of the earlier work were highlighted. We then considered the initial stress to be a
pre-stress associated with the deformation of an isotropic elastic material from a stress-free reference configuration and made
connections with the analysis from the preceding sections.  Specifically, we considered a pure dilatational
deformation and a deformation corresponding to simple tension.  We then discussed the specialization of second-order elasticity
in detail and collated various contributions from the earlier literature that date back to the work of \cite{Bril25}, with
particular reference to expressions for longitudinal and transverse wave speeds, again showing how the results are captured
within the general framework herein.

%%%%%%%%%%%%%%%%%%%%%%%%%%%%%%%%%%%%%%%%%%

\subsection*{Acknowledgement}

This work was in part supported by an International Joint Project grant from the Royal Society of London.

%%%%%%%%%%%%%%%%%%%%%%%%%%%%%%%%%%%%%%%%%%

%%%%%%%%%%%%%%%%%%%%%%%%%%%%%%%%%%%%%%%%%%

\renewcommand\thesection{\Alph{section}}
\setcounter{section}{0}
\renewcommand{\theequation}{A.\arabic{equation}}
\setcounter{equation}{0}

%%%%%%%%%%%%%%%%%%%%%%%%%%%%%%%%%%%%%%%%%%

\section*{Appendix A. Non-principal waves in a solid under initial uniaxial stress \label{AppendixA}}

%%%%%%%%%%%%%%%%%%%%%%%%%%%%%%%%%%%%%%%%%%

Here we complete the analysis of Section \ref{uniaxial-propagation} by considering the case of non-principal wave propagation,
for which the direction of propagation $\mathbf{n}$ and the direction of uniaxial stress $\mathbf{a}$ are neither parallel nor
orthogonal. The constants $A$, $B$, $C$, $D$ are as defined in \eqref{ABCD}.

By solving the equation \eqref{secular} with $\mathbf{Q}(\mathbf{n})$ given by \eqref{Quniaxial} we find that the wave speeds are
given by
\begin{equation}
\rho v^2=A,\quad (A-\rho v^2)^2+[B+C+2D(\mathbf{n}\cdot\mathbf{a})](A-\rho v^2)+(BC-D^2)[1-(\mathbf{n}\cdot\mathbf{a})^2]=0.\label{speeds-uniaxial}
\end{equation}
We therefore consider the cases $\rho v^2=A$ and $\rho v^2\neq A$ separately.

\paragraph{Case 1: $\rho v^2=A.$}

The propagation condition \eqref{prop-cond-comp} yields
\begin{equation}
[B(\mathbf{m}\cdot\mathbf{a})+D(\mathbf{m}\cdot\mathbf{n})]\mathbf{a}+[C(\mathbf{m}\cdot\mathbf{n})
+D(\mathbf{m}\cdot\mathbf{a})]\mathbf{n}=\mathbf{0}.
\end{equation}
If $C(\mathbf{m}\cdot\mathbf{n})+D(\mathbf{m}\cdot\mathbf{a})\neq 0$ then $\mathbf{n}=\pm\mathbf{a}$ and hence $(B+C\pm
2D)(\mathbf{m}\cdot\mathbf{a})=0$.  The case $\mathbf{m}\cdot\mathbf{a}=0$ was covered in Section \ref{uniaxial-propagation} ,
but there is now an additional possibility, that $B+C\pm 2D=0$.  Both these options lead to the same result, which, on
substitution from \eqref{ABCD}, is written
\begin{equation}
\alpha_1+\alpha_2+(3\beta_1+2\beta_2)\tau+(3\gamma_1+2\gamma_2+\beta_3)\tau^2+2\gamma_3\tau^3+\gamma_4\tau^4=0.\label{specialA3}
\end{equation}
There is no restriction on the direction of polarization $\mathbf{m}$. Note that for the specialization \eqref{biot1} this yields
$\tau=-2(\lambda_0+\mu_0)$ and $A=-\lambda_0$ and for several of the values of $\lambda_0$ listed in Table \ref{table3} there is no real wave speed in this case.

Next, consider the possibility that $\mathbf{n}\neq \pm\mathbf{a}$.  Then, if $C(\mathbf{m}\cdot\mathbf{n})
+D(\mathbf{m}\cdot\mathbf{a})= 0$ it follows that also $B(\mathbf{m}\cdot\mathbf{a})+D(\mathbf{m}\cdot\mathbf{n})= 0$. By
combining these we deduce that
\begin{equation}
(BC-D^2)(\mathbf{m}\cdot\mathbf{a})= 0,\quad (BC-D^2)(\mathbf{m}\cdot\mathbf{n})= 0
\end{equation}
provided $C\neq 0$, $D\neq 0$.  Then, if $BC-D^2\neq 0$ we must have $\mathbf{m}\cdot\mathbf{a}=0$ and
$\mathbf{m}\cdot\mathbf{n}=0$ and $\mathbf{m}$ is normal to the plane of $\mathbf{a}$ and $\mathbf{n}$.  Thus, a transverse wave
exists for any direction of propagation.  On the other hand, if $BC-D^2=0$ then $\mathbf{n}$ is determined from the equation
\begin{multline}
BC-D^2\equiv (\alpha_1+ \alpha_2)\{\beta_1\tau+\gamma_1\tau^2+[\beta_3+2\gamma_3\tau+\gamma_4\tau^2]\tau^2(\mathbf{n}\cdot\mathbf{a})^2\}\\[1ex]
-[\beta_1+\beta_2+(\gamma_1+\gamma_2)\tau]^2\tau^2(\mathbf{n}\cdot\mathbf{a})^2=0.\label{BC-D2}
\end{multline}
Since we are considering the case $\mathbf{n}\neq\pm\mathbf{a}$ and $\mathbf{n}\cdot\mathbf{a}\neq 0$, possible directions
$\mathbf{n}$ generate a cone with axis $\mathbf{a}$, provided $|\mathbf{n}\cdot\mathbf{a}|<1$.  For each such $\mathbf{n}$,
$\mathbf{m}$ must satisfy $\mathbf{m}\cdot(B\mathbf{a}+D\mathbf{n})=0$.   Note that in the linear approximation \eqref{BC-D2}
cannot hold unless $\beta_1(0)=0$, in which case $\mathbf{n}$ is unrestricted and $A=\mu_0+\alpha_1'(0)\tau+\tau
(\mathbf{n}\cdot\mathbf{a})^2$.

Other special cases are as follows:  if $B\neq 0$, $C\neq 0$, $D=0$ then $\mathbf{m}\cdot\mathbf{n}=0$,
$\mathbf{m}\cdot\mathbf{a}=0$ and either $\mathbf{n}\cdot\mathbf{a}=0$ or $\beta_1+\beta_2+(\gamma_1+\gamma_2)\tau=0$; if $B=0$,
$C\neq 0$, $D=0$ then $\mathbf{m}\cdot\mathbf{n}=0$ and either $\mathbf{n}\cdot\mathbf{a}=0$ or
$\beta_1+\beta_2+(\gamma_1+\gamma_2)\tau=0$.  In the latter, if $B=0$ and $\mathbf{n}\cdot\mathbf{a}=0$ then
$\beta_1+\gamma_1\tau=0$ (assuming, of course,  $\tau\neq 0$), while if $B=0$ and $\mathbf{n}\cdot\mathbf{a}\neq 0$ then
$(\mathbf{n}\cdot\mathbf{a})^2$ is determined from $B=0$.  Finally, we note that if $C=D=0$ then there are four possibilities:
(i) $\mathbf{n}\cdot\mathbf{a}=0$ and $B=0$ and hence $\beta_1+\gamma_1\tau=0$ --- there is no restriction on $\mathbf{m}$; (ii)
$\mathbf{n}\cdot\mathbf{a}=0$ and $\mathbf{m}\cdot\mathbf{a}=0$ --- this is captured by the discussion around
\eqref{orthogonal-case}; (iii) $\beta_1+\beta_2+(\gamma_1+\gamma_2)\tau=0$ and $B=0$, the latter determining
$(\mathbf{n}\cdot\mathbf{a})^2$ --- there is no restriction on $\mathbf{m}$; (iv) $\beta_1+\beta_2+(\gamma_1+\gamma_2)\tau=0$ and
$\mathbf{m}\cdot\mathbf{a}=0$ --- there is no restriction on $\mathbf{n}$.

\paragraph{Case 2: $\rho v^2\neq A.$}

Now, from the propagation condition \eqref{prop-cond-comp} with the specialization \eqref{Quniaxial} we have
\begin{equation}
(A-\rho v^2)\mathbf{m}+[B(\mathbf{m}\cdot\mathbf{a})+D(\mathbf{m}\cdot\mathbf{n})]\mathbf{a}+[C(\mathbf{m}\cdot\mathbf{n})
+D(\mathbf{m}\cdot\mathbf{a})]\mathbf{n}=\mathbf{0}.\label{neqA}
\end{equation}
Thus, $\mathbf{m},\mathbf{n}, \mathbf{a}$ are coplanar unless either the coefficient of $\mathbf{a}$ or $\mathbf{n}$ vanishes. If
neither of the coefficients vanish then, without loss of generality, we may confine attention to the $(x_1,x_2)$ plane and set
$\mathbf{a}=\mathbf{e}_1$.  Let $(n_1,n_2)$ and $(m_1,m_2)$ be the in-plane components of $\mathbf{n}$ and $\mathbf{m}$,
respectively and set $n_3=m_3=0$.  The propagation condition \eqref{prop-cond-comp} then specializes to the two components
\begin{equation}
Q_{11}m_1+Q_{12}m_2=\rho v^2m_1,\quad Q_{12}m_1+Q_{22}m_2=\rho v^2m_2.\label{propagation2d}
\end{equation}
For a given propagation direction the wave speed is given by one of the two solutions of the quadratic in \eqref{speeds-uniaxial}
and is known explicitly.  Elimination of $\rho v^2$ from \eqref{propagation2d} then gives
\begin{equation}
(Q_{11}-Q_{22})m_1m_2=Q_{12}(m_1^2-m_2^2),\label{propagationm1m2}
\end{equation}
which determines the polarization $\mathbf{m}$.  Let $n_1=\cos\theta,n_2=\sin\theta$ and $m_1=\cos\phi,m_2=\sin\phi$. Then the
above can be rewritten as
\begin{equation}
\tan 2\phi=\frac{C\sin 2\theta +2D\sin\theta}{B+C\cos 2\theta +2D\cos\theta}.\label{tan2phi}
\end{equation}
From this we can immediately recover some of the previous results.  If, for example, $\theta=0$ (propagation in the direction of
initial stress) then $\phi=0$ or $\pi/2$, corresponding to longitudinal and transverse waves, respectively, with wave speeds
given by $\rho v_L^2=A+B+C+2D$ and $\rho v_T^2=A$ (degenerate case).  If $\theta=\pi/2$ (propagation transverse to the initial
stress) then $D=0$ and again $\phi=0$ or $\pi/2$, transverse and longitudinal, respectively, with wave speeds given by $\rho
v_T^2=A+B$ and $\rho v_L^2=A+C$.  There is, however, an additional case not covered previously in which a longitudinal and
transverse wave can propagate.  By setting $\phi=\theta$ in \eqref{tan2phi} and discarding cases already discussed we obtain
$B\cos\theta+D=0$, which leads to (on discarding a factor $\tau\neq 0$)
\begin{equation}
2\beta_1+\beta_2+(2\gamma_1+\gamma_2)\tau+(\beta_3+2\gamma_3\tau+\gamma_4\tau^2)\tau\cos^2\theta=0.\label{2Dtheta}
\end{equation}
If this has a solution (or solutions) $\theta$ for $\cos^2\theta<1$ then a longitudinal wave can propagate in the direction
defined by such an angle (or angles). The corresponding wave speed is given by
\begin{equation}
\rho v_L^2=A+C+D\cos\theta=2\alpha_1+\alpha_2+[1+2\beta_1+\beta_2+(2\gamma_1+\gamma_2)\tau)]\tau\cos^2\theta.\label{a11}
\end{equation}
Equally, by setting $\phi=\theta +\pi/2$ the same reduction $B\cos\theta+D=0$ is obtained and a transverse wave can accompany the
longitudinal wave and has wave speed which, on use of \eqref{2Dtheta}, can be written
\begin{equation}
\rho v_T^2=A+B+D\cos\theta=\alpha_1+\beta_1\tau+\gamma_1\tau^2+\tau\cos^2\theta
-[2\beta_1+\beta_2+(2\gamma_1+\gamma_2)\tau]\tau\sin^2\theta.\label{vTAnot}
\end{equation}
For the specialization \eqref{biot1} equation \eqref{2Dtheta} is satisfied and equations \eqref{a11} and \eqref{vTAnot} reduce to
$\rho v_L^2=\lambda_0+2\mu_0+\tau\cos^2\theta$ and $\rho v_T^2=\mu_0-\frac{1}{2}\tau+\tau\cos^2\theta$, respectively. Formulas in Section 6.4 are recovered by taking $\theta=0$ and $\theta=\frac{1}{2}\pi$.

With reference to \eqref{neqA} we now consider the special cases corresponding to vanishing of one or other of the coefficients
of $\mathbf{a}$ and $\mathbf{n}$. These are $B(\mathbf{m}\cdot\mathbf{a})+D(\mathbf{m}\cdot\mathbf{n})=0$ with
$C(\mathbf{m}\cdot\mathbf{n})+D(\mathbf{m}\cdot\mathbf{a})\neq 0$ and
$B(\mathbf{m}\cdot\mathbf{a})+D(\mathbf{m}\cdot\mathbf{n})\neq 0$ with
$C(\mathbf{m}\cdot\mathbf{n})+D(\mathbf{m}\cdot\mathbf{a})= 0$.  The first of these corresponds to the case $B\cos\theta +D=0$
just considered, while for the second $\mathbf{m}$ is aligned with the direction of uniaxial initial stress and $C\cos\theta
+D=0$, which yields the nontrivial result
\begin{equation}
\alpha_1+\alpha_2+[\beta_1+\beta_2+(\gamma_1+\gamma_2)\tau]\tau=0.\label{special2}
\end{equation}
This puts no restriction on the direction of propagation $\mathbf{n}$ and the wave speed is given by
\begin{equation}
\rho v^2=A+B+D\cos\theta
\end{equation}
as in \eqref{vTAnot}, but the wave is not necessarily transverse in this case.  For the special case \eqref{biot1} the condition
\eqref{special2} yields $\tau=-2(\lambda_0+\mu_0)$, as for \eqref{specialA3}, and $\rho v^2$ specializes to $\mu_0+(\lambda_0+\mu_0)\sin^2\theta$.

%%%%%%%%%%%%%%%

%%%%%%%%%%%%%%%%%%%%%%%%

\end{document}